\documentclass[12pt,preprint,twoside]{article} 
 \usepackage{dsfont}
 \usepackage{amsmath}
 \usepackage{amsthm}
 \usepackage{txfonts,bm}
 \usepackage[sc,bf]{caption}
 \usepackage{latexsym}
 \usepackage{rotating}
 \usepackage{fancyhdr}
 \usepackage{fancyheadings}
 \RequirePackage[OT1]{fontenc}
 \RequirePackage{amsthm,amsmath}
 \RequirePackage[numbers]{natbib}
 \usepackage[colorlinks,citecolor=blue,urlcolor=blue,linkcolor=blue]{hyperref}

 \usepackage{titlesec}
 \titleformat{\section}{\large\bfseries}{\thesection}{1em}{}
 \titleformat{\subsection}{\normalsize\bfseries}{\thesubsection}{1em}{}
 \titleformat{\subsubsection}{\normalsize\it}{{\rm \thesubsubsection}}{1em}{\vspace{-.7em}}

 \newtheoremstyle{mytheo}
                {1ex}
                {0.4ex}
                {\rm}
                {}
                {\bfseries\scshape }
                {.}
                {0.5ex}
                {}

 \theoremstyle{mytheo}

 \newtheorem{theo}{Theorem}[section]

 \newtheorem{cor}{Corollary}[section]
 \newtheorem{defi}{Definition}[section]
 \newtheorem{exam}{Example}[section]
 \newtheorem{lem}{Lemma}[section]
 \newtheorem{prop}{Proposition}[section]
 \newtheorem{rem}{Remark}[section]

 \newenvironment{pr}[1]{\noindent{\bf {#1}:}}{}
 
 \newcommand{\law}{\stackrel{\mbox{\tiny d}}{=}}

 \numberwithin{equation}{section}

 \newcommand{\be}{\begin{equation}}
 \newcommand{\ee}{\end{equation}}
 \newcommand{\E}{\operatorname{\mathds{E}}}
 \renewcommand{\Pr}{\operatorname{\mathds{P}}}

  \newcommand{\RR}{\mathds{R}}

 \title{\Large\bf On sequences of
 expected maxima and expected
 ranges\textcolor[rgb]{0,0,1}{\footnote{Work partially supported
 by the National and Kapodistrian University of
 Athens' Research fund under Grant
 70/4/5637.}}\vspace*{-.6em}}

 \author{\large
 Nickos
 Papadatos
 {\footnote{
 {e-mail:}\
 \textcolor[rgb]{0.98,0.00,0.00}{npapadat@math.uoa.gr}, {url:}\
 \textcolor[rgb]{0.98,0.00,0.00}{users.uoa.gr/$\sim$npapadat/}}}}
 
 \vspace{-1em}

 \date{\small\it
 \begin{tabular}{r@{\hspace{0ex}}l}
 & Department of Mathematics, Section of Statistics and Operational
 Research,
 \\
 [-.3ex]
 &
 National and Kapodistrian University of Athens,
 Panepistemiopolis, 157 84 Athens, Greece.
 \end{tabular}
 }
 \vspace{-1em}

\begin{document}

 \maketitle
 \vspace*{-2em}

 \thispagestyle{empty}

 \begin{abstract}
 \noindent
 We investigate conditions in order to decide whether a given
 sequence of real numbers represents expected maxima or
 expected ranges. The main result provides a novel
 necessary and sufficient condition,
 relating an expected maxima sequence
 to a translation of a Bernstein function
 through its L\'{e}vy-Khintchine representation.
 \end{abstract}
 {\footnotesize {\it MSC}:  Primary 62G30; 60E05; Secondary 44A60.
 \vspace*{-.7ex}
 Bernstein functions, L\'{e}vy-Khintchine representation,
 order statistics.
 }
 \vspace*{-1em}

 \section{Introduction}
 \setcounter{equation}{0}
 \label{sec.1}

 Let $X$ be an integrable random variable (r.v.)\ and suppose that
 $X_{1:k}\leq\cdots \leq X_{k:k}$ are the order statistics arising from
 $k$ independent copies of $X$.
 Based solely on the expected values of order statistics,
 \[
 \mu_{i:k}=\E X_{i:k}, \ \ \ i=1,2,\ldots,k, \ \ k= 1,2,\ldots,
 \]
 Hoeffding (1953) constructed a sequence of r.v.'s $X_k$
 that converges weakly to $X$, and thus, characterized
 the distribution function (d.f.)\ $F$ of $X$ through
 the triangular array  $\mu_{i:k}$. Since each $\mu_{i:k}$ is a
 linear function of $\mu_{i:i}$, $1\leq i\leq k$
 (see Arnold {\it et al.} (1992), p.\ 112, or David and Nagaraja (2003),
 p.\ 45),
 it follows at once that the sequence $\{\mu_k\}_{k=1}^\infty$  of the
 expected maxima $\mu_k=\mu_{k:k}$ uniquely determines the d.f.
 Hill and Spruill (1994), using a theorem of M\"untz (1914),
 improved this result by showing that
 $F$ is characterized by any subsequence $\{\mu_{k(j)}\}_{j=1}^{\infty}$
 with $\sum_{j=1}^{\infty}1/k(j)=\infty$.

 Moreover,  Hill and Spruill (1994, 2000) proved the following
 continuity result:
 \begin{theo}
 \label{theo.1.1}
 Let $\{X_n\}_{n=1}^\infty$ be a sequence of integrable r.v.'s,
 $\{\mu_k\}_{k=1}^\infty$ a sequence of real numbers,
 and write $\mu_k(X_n)$ for the expected maxima of $k$ iid copies of $X_n$.
 If $\mu_k(X_n)\to \mu_k$ as $n\to \infty$ for all $k\geq 1$
 then the following are equivalent.
 \\
 {\rm (i)} There exists an integrable r.v.\ $X$ such that
 $X_n\to_d X$ as $n\to\infty$ {\rm ($\to_d$ denotes
 weak convergence)}
 and $\mu_k(X)= \mu_k$ for all $k\geq 1$.
 \\
 {\rm (ii)}
 $\mu_k=o(k)$ and
 $
 \sum_{j=1}^k (-1)^j {k\choose j} \mu_j=o(k)$
 as $k\to\infty$.
 \end{theo}

 To conclude weak convergence based on this result, it is
 helpful to recognize whether
 a given sequence
 $\{\mu_k\}_{k=1}^\infty$
 represents expected
 maxima of some r.v.
 This question received its own interest, going
 back to Kadane (1971, 1974),
 Mallows (1973), Huang (1998) and Kolodynski (2000). In the sequel,
 a sequence that represents expected
 maxima of some r.v.\ will be called
 {\it Expected Maxima Sequence} (EMS, for short).
 Kadane (1974) proved that
 a necessary and sufficient condition for EMS
 is that the sequence
 $\{\mu_{k+2}-\mu_{k+1}\}_{k=0}^\infty$
 is the moment sequence of a finite measure in the {\it open} interval
 $(0,1)$; that is, there exists a finite measure $\tau$ in $[0,1]$
 such that
 \vspace{-1ex}
 \be
 \label{1.1}
 \tau(\{0\})=\tau(\{1\})=0 \ \mbox{ and } \
 \mu_{n+2}-\mu_{n+1}=\int_{[0,1]} u^n d\tau(u), \ \
 n=0,1,\ldots
 \ .
 \vspace{-.5ex}
 \ee
 According to famous Hausdorff's (1921) characterization, this is equivalent
 \vspace{-.2ex}
 to
 \[
 (-1)^s \Delta^s (\mu_{k+2}-\mu_{k+1})\geq 0, \ \ s\geq 0, \ k\geq 0,
 \vspace{-.2ex}
 \]
 (cf.\ Huang, 1998),
 where $\Delta$ is the forward difference operator ($\Delta^0 \alpha_k=\alpha_k,
 \
 \Delta^1\alpha_k=\Delta\alpha_k=\alpha_{k+1}-\alpha_k, \
 \Delta^{s+1}=\Delta\Delta^s$),
 {\it plus} conditions on the sequence $\mu_k$ that guarantee
 $\tau(\{0\})=\tau(\{1\})=0$.
 Kolodynski (2000) completed Huang's result,
 proving that the boundary
 conditions on the measure $\tau$
 are equivalent to $\mu_k=o(k)$ and
 $\sum_{j=1}^k (-1)^j {k\choose j} \mu_j
 =o(k)$ as $k\to \infty$. Hence,
 another complete characterization of
 EMS's
 is as follows (see Kolodynski, 2000).
 \vspace{-.6ex}
 \begin{theo}
 \label{theo.1.2}
 A sequence $\{\mu_k\}_{k=1}^{\infty}$ represents the expected maxima
 of a non-degenerate integrable r.v.\ if and only if the following three
 conditions are satisfied.
 \\
 {\rm (i)} $(-1)^{s+1} \Delta^s \mu_k>0$ for all $s\geq 1$ and $k\geq 1$.
 \\
 {\rm (ii)} $\mu_k=o(k)$ as $k\to\infty$.
 \\
 {\rm (iii)} $\sum_{j=1}^k (-1)^j {k\choose j} \mu_j
 =o(k)$ as $k\to \infty$.
 \end{theo}

 The purpose of the present work is to give some more light on these
 necessary and sufficient conditions, noting that
 it is rather difficult to check either Kadane's condition
 (\ref{1.1})
 or
 (i)--(iii) of Theorem \ref{theo.1.2} in practical situations;
 we thus provide a much easier
 sufficient condition (of a different nature)
 in Corollary \ref{cor.3.3}. In Section
 \ref{sec.2}
 we present an alternative proof of Theorem \ref{theo.1.2}; the
 interest in this proof lies in its constructive part
 (see Remark \ref{rem.2.1}(b)).

 Section \ref{sec.3} contains the main results,
 Theorems \ref{theo.3.1} and \ref{theo.3.2},
 with illustrative examples
 indicating their usefulness. The main result of Theorem
 \ref{theo.3.1} characterizes the EMS's
 using a novel method that
 relates any such sequence
 to a translation of a suitable Bernstein function
 through its L\'{e}vy-Khintchine representation.
 Finally, in Section \ref{sec.4}
 we provide similar results concerning
 sequences of expected ranges. Several examples
 are given.
 \vspace*{-1ex}


 \section{A probabilistic proof of Theorem \ref{theo.1.2}}
 \setcounter{equation}{0}
 \label{sec.2}

 For completeness of the presentation we give a probabilistic
 proof that only uses the result
 from Hill and Spruill (see Theorem \ref{theo.1.1},
 above) plus the Hoeffding construction;
 thus, we do not invoke results from the moment problem.

 \begin{pr}{Proof of Theorem \ref{theo.1.2}}
 Assume first that $\mu_k=\E X_{k:k}=\mu_k(X)$ for some integrable
 and non-degenerate r.v.\ $X$ with d.f.\ $F$. Then
 we have
 \[
 \mu_k=\int_{-\infty}^\infty \big[I(x>0)-F^k(x)\big] \ dx
 \]
 ($I$ denotes an indicator function),
 and thus, $(-1)^{s+1} \Delta^s \mu_k
 = \int_{-\infty}^\infty F^k(x)(1-F(x))^s dx>0$.
 Also,
 \[
 \frac{\mu_k}{k}=\int_{0}^1 u^{k-1} F^{-1}(u) \ du,
 \]
 where $F^{-1}(u)=\inf\{x:F(x)\geq u\}$, $0<u<1$,
 is the left continuous inverse of $F$. Thus, by dominated convergence
 we conclude that $\lim_{k\to\infty}\frac{\mu_k}{k}=0$. Similarly,
 \[
 \lim_{k\to\infty}
 \int_{0}^1 (1-u)^{k-1} F^{-1}(u) \ du=0,
 \]
 and it is easily seen that $\int_{0}^1 (1-u)^{k-1} F^{-1}(u)du
 =-\frac{1}{k}\sum_{j=1}^k (-1)^j {k\choose j} \mu_j$.

 Conversely, assume that (i)--(iii) are satisfied and define the numbers
 \be
 \label{2.1}
 \beta_{i,n}=\frac{n!}{(i-1)!(n-i)!}\sum_{j=0}^{n-i}{n-i\choose j}
 \frac{(-1)^{j}}{i+j}\mu_{i+j}, \ \ 1\leq i\leq n, \ n\geq 1.
 \ee
 It is easily checked that for every $n\geq 2$ and $1\leq i\leq n-1$,
 \[
 \beta_{i+1,n}-\beta_{i,n}={n\choose i} (-1)^{n-i+1} \Delta^{n-i} \mu_i >0.
 \]
 Therefore, we can define the sequence of discrete uniform
 r.v.'s $X_n$
 by
 \[
 \Pr(X_n=\beta_{i,n})=\frac{1}{n}, \ \ 1\leq i\leq n,
 \]
 noting that the support of $X_n$ is the set $\{\beta_{1,n},\ldots,\beta_{n,n}\}$
 with $\beta_{1,n}<\beta_{2,n}<\cdots<\beta_{n,n}$. Fix now $k\geq 1$ and
 set $Z_{n,k}=\max\{X_{n,1},\ldots,X_{n,k}\}$,
 where $X_{n,1},\ldots,X_{n,k}$
 are iid copies of $X_n$. It is clear that
 $\Pr(Z_{n,k}=\beta_{i,n})=(i/n)^k-((i-1)/n)^k$. Thus,
 \begin{eqnarray*}
 \mu_k(X_n)=\E Z_{n,k}
 &=&
 \sum_{i=1}^n \beta_{i,n} \left[ \left(\frac{i}{n}\right)^k
 -\left(\frac{i-1}{n}\right)^k\right]
 \\
 &=&
 \sum_{i=1}^n \left[ \left(\frac{i}{n}\right)^k
 -\left(\frac{i-1}{n}\right)^k\right]
 \frac{n!}{(i-1)!(n-i)!}\sum_{j=0}^{n-i}{n-i\choose j}
 \frac{(-1)^{j}}{i+j}\mu_{i+j}.
 \end{eqnarray*}
 Substituting $s=i+j$ so that $s\in\{1,\ldots,n\}$ and $j=s-i$, we get
 \begin{eqnarray*}
 \mu_k(X_n)
 &=&
 \sum_{s=1}^n {n \choose s}\frac{\mu_s}{n^k} \sum_{i=1}^s
 (-1)^{s-i} {s-1\choose i-1} \left[i^k-(i-1)^k\right]
 \\
 &=&
 \sum_{s=1}^n {n \choose s}\frac{\mu_s}{n^k} \sum_{i=0}^{s-1}
 (-1)^{s-1-i} {s-1\choose i} \left[(i+1)^k-i^k\right]
 \\
 &=&
 \sum_{s=1}^n {n \choose s}\frac{\mu_s}{n^k} \sum_{m=0}^{k-1}{k\choose m}
 \left\{\sum_{i=0}^{s-1}
 (-1)^{s-1-i} {s-1\choose i} i^m\right\},
 \end{eqnarray*}
 where the term $i^m$ should be treated as 1 if $i=m=0$.

 The expression in the curly brackets is a multiple of
 a Stirling number of the second kind;
 see Charalambides (2002), Theorem 8.4 and p.\ 164.
 Despite this, we can assign a simple probabilistic meaning
 to the sum,
 showing that it vanishes whenever $1\leq m<s-1$.
 Indeed, define
 \[
 S(s,m):=\sum_{i=0}^{s-1}
 (-1)^{s-1-i} {s-1\choose i} \ i^m,
 \]
 and consider $m$
 distinct balls and $s-1$ distinct cells ($s\geq 2, m\geq 1$).
 If we put the balls
 into the cells at random, then the probability that every cell is
 occupied by at least one ball is given by the inclusion-exclusion
 principle:
 \[
 p(s,m):=\Pr(\mbox{every cell contains at least one ball})=\sum_{i=0}^{s-1} (-1)^i
 {s-1\choose i} \left(\frac{s-1-i}{s-1}\right)^m.
 \]
 Hence,
 \[
 p(s,m)=\frac{1}{(s-1)^m}\sum_{i=0}^{s-1} (-1)^{s-1-i} {s-1\choose i} i^m
 =\frac{1}{(s-1)^m}S(s,m).
 \]
 Since the probability $p(s,m)$ is obviously zero whenever $1\leq m<s-1$,
 we conclude that
 $S(s,m)=0$ for $s\geq 3$ and $m=1,\ldots,s-2$.
  In other words, and since $S(s,0)=0$ for $s\geq 2$, we can write
  $S(s,m)=S(s,m)I(s\leq m+1)$, $m\geq 0$, $s\geq 2$,
  where $I$ stands for the indicator function.
 Moreover, since $p(s,s-1)=\frac{(s-1)!}{(s-1)^{s-1}}$ (for $s\geq 2$),
 and $S(1,0)=1$ by convention, we also have
 \[
 S(s,s-1)=(s-1)! \ \mbox{ and } \ S(s,0)=I(s=1), \  \ s\geq 1.
 \]
 Therefore,
 \[
 S(s,m)=S(s,m)I(s\leq m+1), \ \ m\geq 0, \ s\geq 1.
 \]
 Using this observation we see that for $n\geq k$,
 \begin{eqnarray*}
 \mu_k(X_n)
 &=&
 \sum_{s=1}^n  \left(\sum_{m=0}^{k-1}{k\choose m}
 S(s,m)I(s\leq m+1)\right) {n \choose s}\frac{\mu_s}{n^k}
 \\
 &=&
 \sum_{s=1}^k  \left(\sum_{m=s-1}^{k-1}{k\choose m}
 S(s,m)\right) {n \choose s}\frac{\mu_s}{n^k},
 \end{eqnarray*}
 because for $s>k$ we have $I(s\leq m+1)=0$ for all $m=0,\ldots,k-1$.
 Hence,
 \[
 \lim_{n\to \infty} \mu_k(X_n)
 =
 \sum_{s=1}^k \left( \sum_{m=s-1}^{k-1}{k\choose m}
 S(s,m)\right)\lim_{n\to\infty}
 {n \choose s}\frac{\mu_s}{n^k}.
 \]
 Clearly, $\lim_{n\to\infty}
 {n \choose s}\frac{\mu_s}{n^k}=0$ for $s<k$. Thus, only the last term
 ($s=k$) survives, obtaining
 \[
 \lim_{n\to \infty} \mu_k(X_n)
 =
 {k\choose k-1}
 S(k,k-1)\lim_{n\to\infty}
 {n \choose k}\frac{\mu_k}{n^k}=k S(k,k-1) \frac{\mu_k}{k!}=\mu_k.
 \]
 Since $\mu_k(X_n)\to \mu_k$ as $n\to\infty$ for all $k\geq 1$ and, by assumption,
 (ii) and (iii) are satisfied, it follows from Theorem
 \ref{theo.1.1}
 that there exists an integrable $X$
 such that $X_n\to_d X$ and $\mu_k(X)=\mu_k$ for all $k$, completing the proof.
 $\Box$
 \end{pr}

 \begin{rem}
 \label{rem.2.1}
 (a) The construction used in the proof follows the line of Hoeffding (1953);
 the difference here is
 that the numbers $\beta_{i,n}$ in (\ref{2.1})
 are not assumed to be
 expectations
 of (some)
 order statistics.
 \smallskip
 \\
 (b) The proof shows that, under (i), we can always construct a
 sequence $X_n$ such that $\mu_k(X_n)\to \mu_k$ for all $k\geq 1$. However,
 without (ii)
 and (iii) it is possible
 that $X_n\to_d Y$ with $\mu_k(Y)\neq \mu_k$;
 see the examples given in Kolodynski (2000) and in Hill and Spruill (1994).
 \end{rem}

 \begin{exam}
 \label{exam.2.1}
 Let $\mu_k=k-\frac{1}{k+1}$. Then,
 the values $m_k=\mu_{k+2}-\mu_{k+1}=1+\frac{1}{(k+2)(k+3)}$
 correspond to the moments of a finite measure in the interval $[0,1]$.
 More specifically,  one can verify that
 $m_k=\frac{7}{6}\E Y^k$ where
 $F_Y=\frac{6}{7}F_1+\frac{1}{7}F_2$ with $F_1$ being the
 degenerate
 d.f.\ at $1$ (the Dirac measure) and $F_2$ is the
 d.f.\
 of a Beta$(2,2)$ r.v.\ with density $f_2(y)=6y(1-y)$, $0<y<1$.
 Also, a direct calculation using Newton's formula
 shows that for $k\geq0$ and $s\geq 1$,
 \begin{eqnarray*}
 \sum_{j=0}^s (-1)^{j+1} {s \choose j} \left( k+j-\frac{1}{k+j+1}\right)
 &=&
 0+s \sum_{j=0}^{s-1} (-1)^{j} {s-1 \choose j} +
 \sum_{j=0}^{s} (-1)^{j} {s \choose j} \int_{0}^1 x^{k+j} dx
 \\
 &=&
 s I(s=1) + \int_{0}^1 x^k (1-x)^s dx.
 \end{eqnarray*}
 Using the above calculation, it is seen that
 \[
 (-1)^{s+1} \Delta^s \mu_k =\sum_{j=0}^s (-1)^{j+1} {s \choose j} \mu_{k+j}
 = I(s=1)+\frac{k! s!}{(k+s+1)!}>0, \ \ \ k\geq 1, \ \ s\geq 1,
 \]
 and  $\sum_{j=1}^k (-1)^j {k\choose j}\mu_j
 =\frac{k}{k+1}-I(k=1)$. Thus,
 $\mu_k$ satisfies (i) and (iii), but it is not an
 EMS
 since it fails to satisfy (ii). After some algebra it can be
 seen that the numbers $\beta_{i,n}$ in (\ref{2.1}) are
 given by $\beta_{i,n}=\frac{i}{n+1}-1+nI(i=n)$ and the sequence
 of discrete uniform r.v.'s $X_n$, constructed in the proof,
 converges weakly to a Uniform$(-1,0)$ r.v.\
 $X$ with $\mu_k(X)=-\frac{1}{k+1}$; thus, as $n\to\infty$,
 $\mu_k(X_n)\to \mu_k$ for all $k\geq 1$ (because (i) is satisfied --
 see Remark \ref{rem.2.1}(b)),
 $X_n\to_d X$ and
 $\mu_k(X)\neq \mu_k$ for all $k$.
 A similar calculation
 reveals that the sequence
 $\widetilde{\mu_k}=\frac{k}{k+1}-I(k=1)=1-\frac{1}{k+1}-I(k=1)$
 satisfies
 \[
 (-1)^{s+1}\Delta^s \widetilde{\mu_k}=I(k=1)+\frac{k!s!}{(k+s+1)!}
 \ \ \
 \mbox{and}
 \ \ \
 \sum_{j=1}^k (-1)^j {k\choose j}\widetilde{\mu_j} =k-\frac{1}{k+1}, \ \
 k\geq 1, \ s\geq 1.
 \]
 Therefore, (i) and (ii) hold but (iii) fails for $\widetilde{\mu_k}$.
 Now, the corresponding r.v.'s $X_n$
 are uniformly distributed over $\big\{\frac{i}{n+1}-nI(i=1)\big\}_{i=1}^n$
 and, as $n\to\infty$,
 $\mu_k(X_n)\to \widetilde{\mu_k}$ for all $k\geq 1$,
 $X_n\to_d X$ which is Uniform$(0,1)$ and, of course,
 $\mu_k(X)=\frac{k}{k+1}\neq \widetilde{\mu_k}$ only for $k=1$;
 cf.\ the example in
 Hill and Spruill (2000).
 Note that
 $\widetilde{\mu_k}$ and $\mu_k$ are dual sequences
 in the sense that if $\mu_k$ were
 the EMS for some r.v.\ $X$ then $\widetilde{\mu_k}$ would be the EMS for $-X$
 and vice-versa;
 see Kolodynski (2000), p.\ 297.
 \end{exam}

 \section{Necessary and sufficient conditions via integral forms}
 \label{sec.3}
 \setcounter{equation}{0}

 Although the problem of characterizing sequences that represent
 expected maxima
 is completely solved by Theorem
 \ref{theo.1.2} (or (\ref{1.1})), it is usually a difficult task to check
 the conditions (i)--(iii)
 (equiv., to verify existence of $\tau$ in (\ref{1.1}))
 for a given sequence,
 e.g., $\mu_k=\sqrt{k}$ or $\mu_k=\log{k}$.
 In this section we seek for a different kind of necessary and
 sufficient conditions, involving
 the notion of {\it integral forms}, according to the following
 definition (see also Definition \ref{def.3.1}, below).
 \begin{defi}
 \label{def.3.2}
 We say that a function $g:[1,\infty)\to\RR$
 admits a generalized integral form (GIF, for short)
 if there exists
 a finite (positive) measure $\mu$ in $(0,\infty)$,
 and
 measurable functions $h$ and $s$, with $h\geq 0$,
 such that
 \be
 \label{3.2}
 \int_{(0,\infty)} h(y)e^{-y}\big(1-e^{-y}\big) \ d\mu(y)<\infty
 \ \ \
 \mbox{and}
 \ \ \
 g(x)=\int_{(0,\infty)} h(y) \big(s(y)-e^{-xy}\big) \ d\mu(y),
 \ \ x\geq 1.
 \ee
 We shall denote by $\cal{G}$ the class of all such functions and
 by $\cal{G}^*$ the subset of $\cal{G}$ that contains
 all nonconstant functions $g\in\cal{G}$;
 (\ref{3.2})
 will be denoted by $g=G_{s}(h;\mu)$.
 In the particular
 case where $h(y)=h_0(y)$, with
 \be
 \label{3.3}
 h_0(y)=\frac{e^y}{1-e^{-y}}, \ \ \ 0<y<\infty,
 \ee
 we say that
 $g$ is written in canonical form, and we denote
 (\ref{3.2}) by
 \smallskip
  $g=G_s ({\mu})\equiv G_{s}(h_0;\mu)$.
 \end{defi}

 Before proceeding to the main result we present
 some auxiliary results.
 \begin{lem}
 \label{lem.3.1}
 Every $g\in \cal{G}$ can be written in canonical form.
 \end{lem}
 \begin{pr}{Proof}
 For $g=G_s(h;\mu)\in\cal{G}$
 we can define the measure $\nu$ by
 \[
 \nu(A)=\int_A e^{-y} \big(1-e^{-y}\big) h(y) \ d\mu(y), \ \ \
 A \mbox{ Borel }, \ A\subseteq (0,\infty).
 \]
 By (\ref{3.2}) $\nu$ is finite,
 since $\int_{(0,\infty)} d\nu(y)=
 \int_{(0,\infty)} e^{-y} (1-e^{-y}) h(y) d\mu(y)<\infty$.
 Thus,
 \[
 g(x)
 =
 \int_{(0,\infty)} h_0(y) \big(s(y)-e^{-xy}\big)
 \Big(e^{-y}\big(1-e^{-y}\big)h(y)\Big) d\mu(y)
 =
 \int_{(0,\infty)} h_0(y) \big(s(y)-e^{-xy}\big)
  d\nu(y)
 \]
 for all $x\geq 1$, showing that $g=G_s(\nu)$.
 $\Box$
 \end{pr}
 \begin{lem}
 \label{lem.3.2}
 Let $g_1=G_{s_1}(\mu_1)$ and $g_2=G_{s_2}(\mu_2)$
 be two functions in $\cal{G}$. Then, the following are
 equivalent:
 \\
 (i) $g_1(k)-g_2(k)=c$ (constant), $k=1,2,\ldots$ \ \ .
 \\
 (ii) $g_1(x)-g_2(x)=c$ (constant) for all $x\geq 1$.
 \\
 (iii) $\mu_1=\mu_2$.
 \end{lem}
 \begin{pr}{Proof} Since the implications
 (iii)$\Rightarrow$(ii)$\Rightarrow$(i)
 are trivial, we show (i)$\Rightarrow$(iii).
 Clearly, (i) implies $g_2(k)-g_2(1)=g_1(k)-g_1(1)$, i.e.,
 \be
 \label{3.4}
 \int_{(0,\infty)} h_0(y) \big(e^{-y}-e^{-ky}\big)
  d\mu_1(y)
 =
 \int_{(0,\infty)} h_0(y) \big(e^{-y}-e^{-ky}\big)
  d\mu_2(y), \ \ k=2,3,\ldots \  .
 \ee
 Consider the measures $\nu_i$
 ($i=1,2$)
 defined by
 $\nu_i\big((0,u]\big)=\mu_i\big([-\log u,\infty)\big)$,
 $0<u<1$. Changing variables $y=-\log u$ in
 (\ref{3.4}), and since $h_0(-\log u)=1/(u(1-u))$, we obtain
  \[
  \int_{(0,1)} (1+u+\cdots+u^{n})
  \
  d\nu_1(u)
 =
 \int_{(0,1)} (1+u+\cdots+u^{n})
 \
 d\nu_2(u),
  \ \ \ n=0,1,\ldots\ .
 \]
 By induction on $n$ it follows that
 the finite measures $\nu_1$ and $\nu_2$ have all
 their moments equal, and since they have bounded supports,
 they are identical; see, e.g., Billingsley (1995), p.\ 388,
 Theorem 30.1. Therefore, for every $y\in(0,\infty)$,
 $\mu_1\big((0,y]\big)=\nu_1\big([e^{-y},1)\big)
 =
 \nu_2\big([e^{-y},1)\big)
 =\mu_2\big((0,y]\big)$, showing that $\mu_1=\mu_2$.
 $\Box$
 \end{pr}

 \begin{cor}
 \label{cor.3.1}
 The measure $\mu$ in the canonical
 form of $g\in \cal{G}$ is unique. In particular,
 $g(x)=G_s(\mu)(x)=0$ if and only if $\mu=0$;
 any non-vanishing constant function
 $g\notin\cal{G}$.
 \end{cor}

 In the following proposition we
 show that every function $g\in \cal{G}^*$
 is a translation of a {\it Bernstein function}. Recall that a non-negative
 function
 $\beta:[0,\infty)\to[0,\infty)$ is called Bernstein if it is continuous
 on $[0,\infty)$, infinitely differentiable in $(0,\infty)$, and
 its $n$-th order derivative $\beta^{(n)}$ satisfies
 $(-1)^{n+1}\beta^{(n)}(x)\geq 0$ ($n=1,2,\ldots$, $x> 0$);
 cf.\ Schilling {\it et al.} (2012), p.\ 21, Definition 3.1
 (in the sequel, the value $\beta(0)$ will be defined
 by continuity as $\beta(0+)$).
 \begin{prop}
 \label{prop.3.1}
 Let $g=G_s(h;\mu)\in\cal{G}^*$. Then $g$ is continuous on $[1,\infty)$,
 infinitely differentiable in $(1,\infty)$, and its $n$-th order
 derivative is given by
 \be
 \label{3.5}
  (-1)^{n+1}g^{(n)}(x)
  =
  \int_{(0,\infty)}
  y^n h(y) e^{-x y}\ d\mu(y)>0, \ \ n=1,2,\ldots, \ x>1.
 \ee
 \end{prop}
 \begin{pr}{Proof}
 Notice that
 the RHS of
 (\ref{3.5}) is strictly positive for all $x>1$,
 because it can be written as
 $\int_{(0,\infty)} y^n h_0(y) e^{-x y}\ d\nu(y)$, where
 $\nu\neq 0$ is the measure in the canonical form of
 $g$; see Lemma \ref{lem.3.1} and Corollary \ref{cor.3.1}.
 Also, the function
 $g$ is continuous at $x=1$ since for $y>0$ and $\epsilon\in(0,1)$,
 $1-e^{-\epsilon y}\leq 1-e^{-y}$. Hence,
 by (\ref{3.2}) and dominated convergence,
 $g(1+\epsilon)-g(1)
 =\int_{(0,\infty)} h(y) e^{-y}\big(1-e^{-\epsilon y}\big)\ d\mu(y) \to 0
 $,
 as
 $\epsilon\searrow 0$.

 Regarding (\ref{3.5}), we see that
 \hspace{-1ex}
 \[
 \frac{\partial^n}{\partial x^n} \Big(h(y) \big(s(y)-e^{-xy}\big)\Big)
 =(-1)^{n+1} y^n h(y) e^{-xy} \ \ \ (n=1,2,\ldots)
 \hspace{-1ex}
 \]
 is continuous in $x>1$ for every fixed $y>0$. Fix
 $\delta>1$. Then, with $\theta=\delta-1>0$,
 \hspace{-1ex}
 \[
 y^n h(y) e^{-xy}\leq h(y) e^{-y}\big(1-e^{-y}\big)
 \frac{y^n e^{-\theta y}}{1-e^{-y}}\leq
 h(y) e^{-y}\big(1-e^{-y}\big)
 \sup_{y>0}\frac{y^ne^{-\theta y}}{1-e^{-y}}, \ \ x>\delta, \ y>0.
 \hspace{-1ex}
 \]
 The (positive) function $t(y)=y^ne^{-\theta y}/\big(1-e^{-y}\big)$
 is bounded:
 \hspace{-1ex}
 \[
 t(y)\leq \frac{y}{1-e^{-y}}\leq \frac{1}{1-e^{-1}},
 \ 0<y\leq 1; \ \ \ \
 t(y)
 \leq \frac{y^n e^{-\theta y}}{1-e^{-1}}
 \leq \frac{\max\{e^{-\theta},(n/\theta)^n e^{-n}\}}{1-e^{-1}},
 \ \ y>1.
 \hspace{-1ex}
 \]
 Thus, choosing, e.g.,
 $C=\max\{1,(n/\theta)^n e^{-n}\}/(1-e^{-1})$,
 we see that
 \hspace{-1ex}
 \[
 \left|\frac{\partial^n}{\partial x^n} \Big(h(y)
 \big(s(y)-e^{-xy}\big)\Big)
 \right|
 =y^n h(y) e^{-xy}\leq
 C h(y) e^{-y} \big(1-e^{-y}\big), \ \ y>0, \ \ x>\delta.
 \hspace{-1ex}
 \]
 Since
 the dominant function
 $K(y)=C h(y) e^{-y} \big(1-e^{-y}\big)$ is
 integrable
 with respect to $\mu$, it is permitted to
 differentiate (\ref{3.2}) under the integral
 sign (see, e.g., Ferguson 1996, p.\ 124),
 obtaining
 (\ref{3.5}) for $x>\delta>1$;
 and since $\delta>1$ is arbitrary, we conclude
 (\ref{3.5}).
  $\Box$
 \smallskip
 \end{pr}

 Proposition \ref{prop.3.1} shows that if $g\in\cal{G}^*$ then
 the function $B(x):=g(x+1)-g(1)$, $x\geq 0$,
 is Bernstein (of a particular form).
 It is known that every Bernstein function $\beta$ can be
 expressed
 by its {\it L\'{e}vy-Khintchine representation} (LKR, for short)
 \be
 \label{3.6}
 \beta(x)=a_0+a_1 x+\int_{(0,\infty)}\big(1-e^{-xy}\big)
 \
 d\nu(y),
 \ \ x\geq 0;
 \ee
 see Schilling {\it et al.} (2012), p.\ 21, Theorem 3.2. Of course
 it is much simpler to verify the converse, i.e., every function
 that is expressed as in (\ref{3.6}) is
 Bernstein (cf.\ the proof of
 Proposition {\ref{prop.3.1}}).
 The triplet $(a_0,a_1;\nu)$ in LKR is uniquely determined by
 $\beta$, the measure $\nu$ satisfies
 $\int_{(0,\infty)}\min\{1,y\}\ d\nu(y)<\infty$,
 and the constants
 $a_0$, $a_1$ are non-negative. Comparing the
 LKR
 of $B$
 with the canonical form of
 $g=G_s(\mu)\in \cal{G}^*$, we see that
 (see (\ref{3.2}))
 \[
 a_0+a_1 x+\int_{(0,\infty)}\big(1-e^{-xy}\big)
 \ d\nu(y)
 =g(x+1)-g(1)
 =
 \int_{(0,\infty)} e^{-y}h_0(y) \big(1-e^{-xy}\big)
 \ d\mu(y), \ \ x\geq0.
 \]
 That is, $a_0=a_1=0$ and $d\nu(y)=e^{-y}h_0(y)d\mu(y)$
 is the
 LKR
 of
 $B(x)=g(x+1)-g(1)$.
 Conversely, if  $\cal{B}^{*}$
 denotes the
 class of Bernstein functions with
 LKR triplet $(0,0;\nu)$, $\nu\neq 0$, it is not difficult
 to show that $g(x+1)-g(1)\in \cal{B}^*$ implies
 $g\in\cal{G}^{*}$.
 Hence, $g\in\cal{G}^{*}$
 if and only if
 $B\in \cal{B}^*$, and we conclude
 the following:

 \begin{prop}
 \label{prop.3.2}
 A function $g:[1,\infty)\to\RR$ belongs to
 $\cal{G}^*$ if and only if
 $B(x):=g(x+1)-g(1)$, $x\geq 0$, is a Bernstein function that
 admits
 a L\'{e}vy-Khintchine representation of the form
 (\ref{3.6})
 with $a_0=a_1=0$, $\nu\neq 0$.
 \end{prop}

 We are now in a position to state and prove the
 main result.
 \begin{theo}
 \label{theo.3.1}
 For a real sequence $\{\mu_k\}_{k=1}^{\infty}$ the following are equivalent:
 \\
 {\rm (i)} There exists a non-degenerate integrable r.v.\
 $X$ such that $\mu_k(X)=\mu_k$ for $k=1,2,\ldots$ \ .
 \\
 {\rm (ii)}
 The sequence
 $\{\mu_k\}_{k=1}^{\infty}$
 is the restriction to the natural numbers
 of a function $g\in\cal{G}^*$
 (for $\cal{G^*}$ see Definition \ref{def.3.2}),
 i.e., $\mu_k=g(k)$, $k=1,2,\ldots$
 \\
 {\rm (iii)} There exists a Bernstein function $B$ with
 L\'{e}vy-Khintchine triplet $(0,0;\nu)$, $\nu\neq 0$
 (see (\ref{3.6})),
 such that $\mu_k=\mu_1+B(k-1)$, $k=1,2,\ldots$ \ .

 If one of (i), (ii), (iii) is fulfilled by
 $\{\mu_k\}_{k=1}^{\infty}$, then the function
 $g\in\cal{G}^*$ in (ii) is unique,
 and admits the representation
 \be
 \label{3.7}
 g(x)=
 \int_{(0,\infty)} \frac{\lambda e^y}{1-e^{-y}}
 \left(\frac{\mu_1}{\lambda} e^{-y} (1-e^{-y})
 +e^{-y}-e^{-xy}\right)  d F_Y(y), \ \ x\geq 1,
 \ee
 where $\lambda=\mu_2-\mu_1$, $F_Y$ is the d.f.\
 of the r.v.\ $Y=-\log F(V)$,
 $F$ is the d.f.\ of $X$, and the r.v.\ $V$
 has density
 \[
 f_V(x)=\frac{1}{\lambda} F(x)(1-F(x)), \ \
 \ \
 -\infty<x<\infty;
 \]
 the Bernstein function $B$ in (iii), which is also unique,
 is related to $g$ by $B(x)=g(x+1)-\mu_1$, $x\geq0$.

 \end{theo}
 \begin{pr}{Proof} (ii)$\Rightarrow$(i).
 Suppose that $\mu_k=g(k)$, $k=1,2,\ldots$, for some
 $g=G_{s}(h;\mu)\in\cal{G}^*$. It suffices to verify
 conditions (i)--(iii) of Theorem \ref{theo.1.2} for $\mu_k$.
 From (\ref{3.5}), $g'(x)>0$  for $x>1$.
 Hence,
 by monotone convergence and by continuity of $g$ at
 $1+$,
 \be
 \label{3.8}
 \int_{1}^{x} g'(t) \  dt =
 \lim_{\epsilon\searrow 0}
 \int_{1+\epsilon}^{x} g'(t) \ dt=
 \lim_{\epsilon\searrow 0} \big[g(x)-g(1+\epsilon)\big]=
 g(x)-g(1), \ \ x> 1.
 \ee
 It should be noted that differentiability of $g$ in $(1,\infty)$ plus
 continuity at $1$ are not sufficient for concluding (\ref{3.8}), as
 the example $g(x)=(x-1)\sin(1/(x-1))$ shows.
 Now, by induction on $s$ (and by using (\ref{3.8}) when
 $k=1$), it is easily
 seen that
 \be
 \label{3.9}
 \sum_{j=0}^s (-1)^{s-j} {s\choose j} g(k+j)=\int_k^{k+1}
 \int_{t_1}^{t_1+1}\cdots\int_{t_{s-1}}^{t_{s-1}+1} g^{(s)}(t_s)
 \ dt_s\ldots
 dt_2 dt_1, \ \ s\geq 1, k\geq 1.
 \ee
 Therefore, since $\mu_{k+j}=g(k+j)$,
 \[
 (-1)^{s+1}\Delta^s \mu_{k}=
 \sum_{j=0}^s (-1)^{j+1} {s\choose j} g(k+j)
 =\int_k^{k+1}
 \int_{t_1}^{t_1+1}\cdots\int_{t_{s-1}}^{t_{s-1}+1} (-1)^{s+1}g^{(s)}(t_s)
 \
 dt_s\ldots
 dt_2 dt_1;
 \]
 the last expression verifies condition (i) of Theorem
 \ref{theo.1.2},
 because the integrand
 is strictly positive (see (\ref{3.5})). Condition (ii) of Theorem
 \ref{theo.1.2} is simply deduced from
 dominated convergence since $(1-e^{-ky})/k\leq 1-e^{-y}$
 and, obviously, $(1-e^{-ky})/k\to 0$ as $k\to\infty$. Hence,
 \[
 \lim_{k\to\infty} \frac{\mu_k}{k}=\lim_{k\to\infty}
 \frac{\mu_{k+1}-\mu_1}{k}=
 \lim_{k\to\infty} \int_{(0,\infty)} e^{-y} h(y)
 \left(\frac{1-e^{-ky}}{k}\right)  d\mu(y)=0.
 \]
 Set now $\nu_k=\sum_{j=1}^k (-1)^j {k\choose j}\mu_j$, so that $\nu_1=-\mu_1$.
 It is not hard to check that $\nu_{s+1}-\nu_s=(-1)^{s+1} \Delta^s \mu_1>0$, where
 $\Delta^s \mu_1=\sum_{j=0}^s (-1)^{s-j} {s\choose j}\mu_{j+1}$.
 Defining $y_s:=(-1)^{s+1}\Delta^s \mu_1>0$, we have
 \[
 \nu_k=-\mu_1+\sum_{s=1}^{k-1} (-1)^{s+1}\Delta^s \mu_{1}=-\mu_1
 +\sum_{s=1}^{k-1} y_s.
 \]
 If it can be shown that $\lim_{k\to\infty} y_k=0$ then it will follow
 that
 \[
 \lim_{k\to\infty} \frac{\nu_k}{k} =\lim_{k\to\infty}
 \frac{y_1+\cdots+y_{k-1}}{k}=0,
 \]
 which means that the sequence $\mu_k$ satisfies condition (iii) of Theorem
 \ref{theo.1.2}. Due to (\ref{3.2}),
 \begin{eqnarray*}
 y_k=\sum_{j=0}^k (-1)^{j+1} {k\choose j}g(j+1)
 & = &
 \sum_{j=0}^k (-1)^{j+1} {k\choose j}
 \int_{(0,\infty)}
 h(y) \Big(s(y)-e^{-(j+1)y}\Big)\ d\mu(y)
 \\
 &=&
 \int_{(0,\infty)}
 h(y) e^{-y} (1-e^{-y}\big)^k \ d\mu(y) \to 0, \ \
 \mbox{as} \ k\to\infty,
 \end{eqnarray*}
 by dominated convergence.

 (i)$\Rightarrow$(ii). Let $F$
 be the d.f.\ of $X$, and set
 $\alpha=\inf\{x:F(x)>0\}$, $\omega=\sup\{x:F(x)<1\}$.
 By the assumption that $X$ is non-degenerate it follows
 that $-\infty\leq \alpha<\omega\leq +\infty$, and the
 open interval $(\alpha,\omega)$ has strictly positive (or infinite)
 length. We define the family of
 d.f.'s
 $\{F^t, t\geq 1\}$,
 and let us denote by $X_t$ a generic r.v.\ with
 d.f.\
 $F^t$, so that $X_1=X$. Since $X$ is integrable, the same is true
 for each $X_t$. Indeed, $F^{t}(x)\leq F(x)$ and $1-F^t(x)\leq t(1-F(x))$
 for all $x\in\RR$ and $t\geq 1$, showing that
 \begin{eqnarray*}
 \E X_t^{-}
 &=&
 \int_{-\infty}^0 F^t(x) \ dx\leq
 \int_{-\infty}^0 F(x) \ dx <\infty,
 \\
 \E X_t^{+}
 &=&
 \int_{0}^{\infty} (1-F^t(x)) \ dx\leq
 t \int_0^{\infty} (1-F(x)) \ dx <\infty,
 \end{eqnarray*}
 where $X^+=\max\{X,0\}$, $X^-=\max\{-X,0\}$,
 denotes, resp., the positive and negative part of
 any r.v.\ $X$. This enables us to define the function
 $g:[1,\infty)\to\RR$ by
 \[
 g(t):=\E X_t=\int_{-\infty}^{\infty} [I(x>0)-F^t(x)] \ dx, \ \ t\geq 1;
 \]
 by definition, $g(k)=\mu_k$ for $k=1,2,\ldots$ .
 For $t\in[1,\infty)$
 write
 \begin{eqnarray}
 \nonumber
 \hspace*{-5ex}
 g(t)-g(1)
 \hspace*{-1ex}&=&\hspace*{-1ex}
 \int_{\alpha}^{\omega} [F(x)-F^t(x)] \ dx
 =
 \int_{\alpha}^{\omega} F(x)(1-F(x)) \frac{F(x)-F^t(x)}{F(x)(1-F(x))} \ dx
 \\
 \hspace*{-5ex}
 \hspace*{-1ex}&=&\hspace*{-1ex}
 \label{3.10}
 \int_{\alpha}^{\omega} F(x)(1-F(x))
 \frac{e^{-\delta(x)}-e^{-t\delta(x)}}{e^{-\delta(x)}(1-e^{-\delta(x)})} \ dx,
 \ \mbox{where $\delta(x)=-\log F(x)$;}
 \end{eqnarray}
 note that $0<F(x)<1$
 for all $x\in (\alpha,\omega)$, so that $\delta(x)>0$. Setting
 $\lambda=\mu_2-\mu_1=g(2)-g(1)=\int_{\alpha}^{\omega} F(x)(1-F(x)) dx >0$,
 we readily see that $f_V(x):=F(x)(1-F(x))/\lambda$ defines a
 probability density
 on $\RR$ with support $(\alpha,\omega)$. Consider an
 r.v.\ $V$ with density $f_V$. Then (\ref{3.10})
 can be rewritten as
 \[
 g(t)-g(1)=\lambda \E \left\{\frac{e^{\delta(V)}}{1-e^{-\delta(V)}}
 \Big(e^{-\delta(V)}-e^{-t\delta(V)}\Big)\right\},
 \ \ t\geq 1,
 \]
 where $\delta(V)=-\log F(V)$ is a strictly positive
 r.v., because $\alpha<V<\omega$
 w.p.\ 1. Setting $Y:=\delta(V)>0$, we get
 \[
 g(t)-g(1)=\lambda \E \left\{\frac{e^{Y}}{1-e^{-Y}}
 \Big(e^{-Y}-e^{-tY}\Big)\right\}
 =\lambda \int_{(0,\infty)} h_0(y) (e^{-y}-e^{-t y}) \ d F_Y(y),
 \ \ t\geq 1,
 \]
 where $h_0(y)=e^y/(1-e^{-y})$ (see (\ref{3.3}))
 and $F_Y$ is the
 d.f.\
 of $Y$. If we introduce the measure $\mu$ defined by
 $\mu(A)=\lambda \Pr(Y\in A)$
 for Borel  $A\subseteq (0,\infty)$,
 the above relation takes the form
 \[
 g(t)-g(1)=\int_{(0,\infty)} h_0(y) (e^{-y}-e^{-t y}) \ d \mu(y),
 \ \ t\geq 1.
 \]
 Moreover, since $h_0(y)>0$,
 \[
 0< \int_{(0,\infty)} h_0(y) e^{-y}(1-e^{-y}) \ d \mu(y)
 =
 \int_{(0,\infty)} d \mu(y)=\mu((0,\infty))=\lambda<\infty.
 \]
 Observing that
 \[
 g(1)=\mu_1=\frac{\mu_1}{\lambda} \int_{(0,\infty)} d \mu(y)
 =\int_{(0,\infty)} h_0(y) \left(\frac{\mu_1}{\lambda} e^{-y} (1-e^{-y})
 \right)
 d \mu(y),
 \]
 we get
 \[
 g(t)=
 g(1) + \big(g(t)-g(1)\big) =
 \int_{(0,\infty)} h_0(y)
 \left(\frac{\mu_1}{\lambda} e^{-y} (1-e^{-y})
 +e^{-y}-e^{-ty}\right)  d \mu(y), \ \ t\geq 1;
 \]
 this shows both (\ref{3.2}) and (\ref{3.7}).

 Finally, the equivalence of (ii) and (iii) follows from Proposition
 \ref{prop.3.2}, and uniqueness
 (of $g$ and $\mu$)
 is evident from Lemma \ref{lem.3.2}.
 %
 %
 $\Box$
 \smallskip
 \end{pr}


 The following definition provides a helpful tool
 in verifying whether a given function $g$ belongs to $\cal{G}^*$.

 \begin{defi}
 \label{def.3.1}
 Let $g:[1,\infty)\to\RR$ be an arbitrary function.
 We say that $g$ admits an integral form (IF, for short)
 if there exist
 measurable functions $h_1:(0,\infty)\to\RR$ and $s:(0,\infty)\to\RR$,
 with $h_1\geq 0$,
 such that
 \begin{eqnarray}
 &&
 \label{3.1a}
 0<\int_{0}^{\infty} h_1(y) e^{-y}\big(1-e^{-y}\big)
  \ dy<\infty
 \\
 &&
 \nonumber
 \hspace{-7ex}\mbox{and}
 \\
 &&
 \label{3.1b}
 g(x)=\int_{0}^{\infty} h_1(y) \big(s(y)-e^{-xy}\big)
 \ dy, \ \ x\geq 1.
 \end{eqnarray}
 We shall denote by $\cal{I}$ the class of all such functions and,
 provided that
 $h_1$ satisfies (\ref{3.1a}),
 the representation (\ref{3.1b}) will be denoted by $g=I_s(h_1)$.
 \end{defi}

 \begin{lem}
 \label{lem.3.3}
 $\cal{I}\subseteq\cal{G}^*$.
 \end{lem}
 \begin{pr}{Proof}
 Assume that $g=I_s(h_1)\in\cal{I}$ and define the (positive)
 measure $\mu$ by
 \[
 \mu\big((0,y]\big)=\int_{0}^y h_1(x) e^{-x}
 \big(1-e^{-x}\big)\ dx, \ \ y> 0.
 \]
 By definition,
 $\mu$ is absolutely continuous with respect to
 Lebesgue measure on $(0,\infty)$,
 with Radon-Nikodym derivative
 \[
 \frac{d \mu(y)}{d y}=h_1(y) e^{-y} \big(1-e^{-y}\big), \ \
 \mbox{for almost all} \ \ y>0.
 \]
 Clearly $\mu$ is finite,
 and
  (\ref{3.1b})
 can be rewritten as
 \[
 g(x)=\int_{0}^{\infty} h_0(y) \big(s(y)-e^{-xy}\big)
 \Big(h_1(y)e^{-y}\big(1-e^{-y}\big)\Big)\ d y
 =\int_{(0,\infty)}
 h_0(y)
 \big(s(y)-e^{-xy}\big)
 \ d \mu(y), \ \ x\geq 1,
 \]
 showing the integral representation in
 (\ref{3.2}).
 Moreover, from (\ref{3.1a}),
 \[
 0<\int_{(0,\infty)} h_0(y) e^{-y}\big(1-e^{-y}\big) \ d \mu(y)
 =
 \int_{(0,\infty)} \ d \mu(y)
 =\int_{0}^{\infty} h_1(y) e^{-y}\ \big(1-e^{-y}\big)\ dy<\infty.
 \]
 Hence, $g=G_{s}(\mu)$ with $\mu\neq 0$.
 $\Box$
 \end{pr}
 \begin{cor}
 \label{cor.3.2}
 The function $h_1$ in the integral representation
 (\ref{3.1b}) of any $g=I_s(h_1)\in\cal{I}$ is (almost
 everywhere) unique.
 \end{cor}
 \begin{pr}{Proof}
 If we express $g=I_s(h_1)\in \cal{G}$
 in its canonical form as $g=G_s(\mu)$ (see Lemma \ref{lem.3.1}),
 then the function $h_1(y)/h_0(y)$ is a
 Radon-Nikodym derivative of
 $\mu$ with respect to  Lebesgue measure.
 The result follows from
 Corollary \ref{cor.3.1} and the fact that
 the Radon-Nikodym derivative is almost everywhere unique.
 $\Box$
 \smallskip
 \end{pr}

 We can now state  the following result which
 provides a sufficient condition
 that is useful
 for most practical situations.
 \begin{cor}
 \label{cor.3.3}
 If a function $g:[1,\infty)\to\RR$
 belongs to $\cal{I}$ (see Definition \ref{def.3.1})
 then
 the sequence
 $\mu_k=g(k)$, $k=1,2,\ldots$,
 represents the expected maxima sequence of an integrable
 non-degenerate random variable.
 \end{cor}
 \begin{pr}{Proof}
 Evident from Theorem \ref{theo.3.1} and Lemma
 \ref{lem.3.3}.
 $\Box$
 \smallskip
 \end{pr}

 If $g=G_{s}(\mu)\in\cal{G}^*$
 (see Definition \ref{def.3.2})
 and the measure $\mu$
 has a Radon-Nikodym derivative $h_\mu$ with
 respect to Lebesgue measure,
 the condition (\ref{3.2})  is equivalent
 to (\ref{3.1a}) and (\ref{3.1b}). Indeed, in this case,
 \[
 g(x)=\int_{(0,\infty)}
 h_0(y)
 \big(s(y)-e^{-xy}\big)
 \ d \mu(y)=
 \int_{0}^{\infty} h_\mu(y) h_0(y) \big(s(y)-e^{-xy}\big)
 \ d y,
 \]
 and it is sufficient to choose $h_1=h_0\cdot h_\mu$.
 Hence, $g=G_s(\mu)\in\cal{I}$ if and only if the measure
 $\mu$ in the canonical form of $g$
 is (non-zero and) absolutely continuous with respect to Lebesgue
 measure.
 However, given an {\it arbitrary} sequence $\mu_k$,
 even if it can be shown that it is an EMS
 (using, e.g., Theorem \ref{theo.1.2}, Theorem \ref{theo.3.1},
 Corollary \ref{cor.3.3},
 or (\ref{1.1})),
 we would like to decide if it corresponds to
 an absolutely continuous r.v.
 We note at this point that the condition $g\in\cal{I}$
 is neither necessary nor sufficient for concluding
 that the EMS $\{g(k)\}_{k=1}^\infty$
 corresponds to a density (see Remark \ref{rem.3.4}, below).
 An interesting exception where this fact can be deduced
 automatically is described by the following definition.
 \begin{defi}
 \label{def.3.3}
 Denote by $\cal{F}$ the subclass of absolutely continuous
 r.v.'s $X$ with interval supports
 $(\alpha,\omega)=(\alpha_X,\omega_X)$,
 $-\infty\leq \alpha<\omega\leq+\infty$,
 having a
 differentiable
 d.f.\
 $F$ in $(\alpha,\omega)$,
 and such that their density $f(x)=F'(x)$
 is strictly positive and continuous in $(\alpha,\omega)$.
 \end{defi}
 \begin{theo}
 \label{theo.3.2}
 For a given sequence $\{\mu_k\}_{k=1}^\infty$ the
 following statements are equivalent:
 \\
 (i) The sequence $\mu_k$ represents an
 expected maxima
 sequence of an integrable r.v.\
 $X\in \cal{F}$.
 \\
 (ii)
 There is an extension
 $g:[1,\infty)\to\RR$ of the sequence $\mu_k$
 {\rm (that is, $\mu_k=g(k)$, $k=1,2,\ldots$)},
 such that $g$
 admits an integral representation
 of the form (\ref{3.1b}), with
 $h_1$
 satisfying (\ref{3.1a}) and, furthermore,
 $h_1$ is strictly positive and continuous
 in $(0,\infty)$.
 \smallskip

 Moreover, if (i) or (ii) holds, then the function $g$
 is unique, and the continuous version of
 $h_1$ in the integral representation
 (\ref{3.1b})
 is uniquely determined by
 \be
 \label{3.13}
 h_1(y)=\frac{e^{-y}}{f(F^{-1}(e^{-y}))}, \ \ 0<y<+\infty,
 \ee
 where $f$ and $F^{-1}$ are, respectively, the density and
 the inverse d.f.\ of the unique r.v.\ $X\in\cal{F}$ with
 expected
 maxima $\mu_k$; any other version $h_2$ is equal to
 $h_1$ almost everywhere in $(0,\infty)$.
 \end{theo}
 \begin{pr}{Proof}
 Assume first that (i) holds, let $F$
 be the d.f.\ of $X$, and set
 $(\alpha,\omega)=\{x:0<F(x)<1\}$.
 Since $X\in\cal{F}$,
 $\lambda:=\mu_2-\mu_1>0$.
 Using (\ref{3.7}) and
 the fact that $V$
 has density
 \[
 f_V(x)=\frac{1}{\lambda} F(x)(1-F(x)), \ \
 \ \
 \alpha<x<\omega,
 \]
 the additional assumption $X\in\cal{F}$
 implies that $Y=-\log F(V)$ has a continuous,
 strictly positive, density
 \[
 f_Y(y)=
 \frac{e^{-2y}\big(1-e^{-y}\big)}{\lambda f(F^{-1}(e^{-y}))},
 \  \  \ \
 0<y<\infty,
 \]
 with $f$ and $F^{-1}$ being, respectively, the derivative and the
 ordinary inverse of the restriction in $(\alpha,\omega)$ of $F$.
 Substituting
 $d F_Y(y)=f_Y(y) dy$ in (\ref{3.7}) we get (ii)
 with $h_1$ as in (\ref{3.13}).

 Assume now that (ii) holds.
 From (\ref{3.1b}),
 \be
 \label{3.14}
 \mu_k-\mu_1=g(k)-g(1)=\int_{0}^{\infty} h_1(y)\big(e^{-y}-e^{-k y}\big)\ dy,
 \ \ \
 k=1,2,\ldots \ .
 \ee
 Also, Corollary \ref{cor.3.3} shows
 that the sequence $\mu_k=g(k)$ is an EMS
 of a unique (non-degenerate) r.v.\ $X$. It
 remains to show that $X\in\cal{F}$, i.e.,
 that its d.f.\ $F$ belongs to $\cal{F}$. To this end,
 define the function
 \be
 \label{3.15}
 G(u):=\left\{
 \begin{array}{ll}
 c_1-\int_{u}^{1/2} \frac{1}{t} h_1(-\log t) \ dt, & 0<u\leq 1/2,
 \\
 & \vspace{-1ex}
 \\
 c_1+\int_{1/2}^{u} \frac{1}{t} h_1(-\log t) \ dt, & 1/2<u<1,
 \end{array}
 \right.
 \ee
 where $c_1$ is a constant to be specified later. By the assumption
 on $h_1$, $G$
 is strictly increasing and differentiable in the interval $(0,1)$.
 Moreover, $G$ is integrable, since by (\ref{3.1a}) and Tonelli's theorem,
 \begin{eqnarray*}
 \int_{0}^1 |G(u)-c_1| \ du
 &=&
 \int_{0}^{1/2} \int_{u}^{1/2}
 \frac{1}{t} h_1(-\log t) \ dt \ du
 +
 \int_{1/2}^{1} \int_{1/2}^{u}
 \frac{1}{t} h_1(-\log t) \ dt \ du
 \\
 &=&
 \int_{0}^{1/2}  h_1(-\log t) \ dt
 +
 \int_{1/2}^{1} \frac{1-t}{t} h_1(-\log t) \ dt
 \\
 &=&
 \int_{\log2}^{\infty}  e^{-y} h_1(y) \ dy
 +
 \int_{0}^{\log2} \big(1-e^{-y}\big) h_1(y) \ dy <\infty.
 \end{eqnarray*}
 Let $U$ be a Uniform$(0,1)$ r.v.\ and define
 the r.v.\ $Y:=G(U)$ with d.f.\ $F_Y$;
 that is, $G=F_Y^{-1}$. Clearly, $\E|Y|=\E|G(U)|<\infty$.
 We can show that $Y\in \cal{F}$. Indeed,
 setting $\alpha_Y:=\lim_{u\searrow 0}G(u)$,
 $\omega_Y:=\lim_{u\nearrow 1}G(u)$, we see that
 $G:(0,1)\to (\alpha_Y,\omega_Y)$ is strictly increasing
 and differentiable, with continuous, strictly positive,
 derivative $G'(u)=h_1(-\log u)/u$. This means that
 its inverse, $G^{-1}=F_Y:(\alpha_Y,\omega_Y)\to (0,1)$,
 has also a continuous, strictly positive, derivative
 $f_Y(y)=F_Y'(y)=1/G'(G^{-1}(y))$. Observe that $F_Y(y)=G^{-1}(y)$
 tends to $0$ as $y$ approaches $\alpha_Y$
 from above, so that, by monotone convergence,
 \[
 \int_{\alpha_Y} ^ {y} f_Y(x) \ dx =
 \lim_{a\searrow \alpha_Y}
 \int_{a} ^ {y} F'_Y(x) \ dx =
 \lim_{a\searrow \alpha_Y}
 [F_Y(y)-F_Y(a)]= F_Y(y), \ \ \ \alpha_Y<y<\omega_Y.
 \]
 Taking limits as $y\nearrow \omega_Y$ in the above relation,
 and using
 again monotone convergence and the fact that
 $G^{-1}(y)$ tends to $1$ as $y\nearrow \omega_Y$, we see that
 \[
 \int_{\alpha_Y}^{\omega_Y} f_Y(x) \ dx =
 \lim_{y\nearrow \omega_Y}
 \int_{\alpha_Y}^{y} f_Y(x) \ dx =
 \lim_{y\nearrow \omega_Y}
 F_Y(y)=
 \lim_{y\nearrow \omega_Y}
 G^{-1}(y)=
  1;
 \]
 hence, $Y\in\cal{F}$. According to the
 implication (i)$\Rightarrow$(ii), the sequence
 $\widetilde{\mu}_k:=\mu_k(Y)$
 admits an extension $g_2:[1,\infty)\to\RR$
 of the form
 \[
 g_2(x)=\int_{0}^{\infty} h_2(y) \big(s_2(y)-e^{-xy}\big) \ dy,
 \ \ \ x\geq 1,
 \]
 such that $h_2$ satisfies (\ref{3.1a}) (with $h_2$ in place of $h_1$)
 and is continuous
 and strictly positive in $(0,\infty)$. Therefore,
 we have
 \be
 \label{3.16}
 \widetilde{\mu}_k-\widetilde{\mu}_1=g_2(k)-g_2(1)=
 \int_{0}^{\infty} h_2(y) \big(e^{-y}-e^{-ky}\big) \ dy,
 \ \ \ k=1,2,\ldots \ .
 \ee
 We can calculate the same quantities directly from $G=F_Y^{-1}$
 as follows:
 \begin{eqnarray*}
 \widetilde{\mu}_k-\widetilde{\mu}_1
 &= &
 \int_{0}^1 k u^{k-1} G(u) \ du - \int_{0}^1  G(u) \ du
 \\
 &=&
 -\int_{0}^{1/2} k u^{k-1}\int_{u}^{1/2}
 \frac{1}{t} h_1(-\log t) \ dt \ du
 +\int_{0}^{1/2} \int_{u}^{1/2}
 \frac{1}{t} h_1(-\log t) \ dt \ du
 \\
 &&
 +
 \int_{1/2}^{1} ku^{k-1}\int_{1/2}^{u}
 \frac{1}{t} h_1(-\log t) \ dt \ du
 -
 \int_{1/2}^{1} \int_{1/2}^{u}
 \frac{1}{t} h_1(-\log t) \ dt \ du.
 \end{eqnarray*}
 Since all integrands in the last four integrals are non-negative,
 we can interchange the order of integration. Thus,
 \begin{eqnarray}
 \nonumber
 \widetilde{\mu}_k-\widetilde{\mu}_1
 &= &
 -\int_{0}^{1/2}
 t^{k-1} h_1(-\log t) \ dt
 +\int_{0}^{1/2}  h_1(-\log t) \ dt
 \\
 \nonumber
 &&
 +
 \int_{1/2}^{1}
 \frac{1-t^k}{t} h_1(-\log t) \ dt
 -
 \int_{1/2}^{1}
 \frac{1-t}{t} h_1(-\log t) \ dt
 \\
 \nonumber
 &= &
 \int_{\log2}^{\infty}
 \big(-e^{-ky}+e^{-y}\big) h_1(y) \ dy
 +
 \int_{0}^{\log2}
 \Big(\big(1-e^{-ky}\big)-\big(1-e^{-y}\big)\Big) h_1(y) \ dy
 \\
 \label{3.17}
 &=&
 \int_{0}^{\infty}
 \big(e^{-y}-e^{-ky}\big) h_1(y) \ dy,
 \ \
 \ \ \
 k=1,2,\ldots \ .
 \end{eqnarray}
 From (\ref{3.14}), (\ref{3.16}) and (\ref{3.17}) we see that
 \[
 g(k)-g(1)=\mu_k-\mu_1
 =\widetilde{\mu}_k-\widetilde{\mu}_1=
 g_2(k)-g_2(1), \ \ \ k=1,2,\ldots \ .
 \]
 Therefore, since $g$ and $g_2$ belong to $\cal{I}\subseteq \cal{G}^{*}$,
 it follows from Lemma \ref{lem.3.2} that
 $g(x)-g_2(x)=\mu_1-\widetilde{\mu}_1$
 (constant),
 $x\geq 1$.
 Choosing
 the constant $c_1$ in (\ref{3.15}) so
 that $\widetilde{\mu}_1=\mu_1$,
 we get $\widetilde{\mu}_k=\mu_k$ for all $k$, which implies
 that $g=g_2$ and $F=F_Y\in\cal{F}$.

 Uniqueness of $g$ and $h_1$ follow immediately from
 Theorem \ref{theo.3.1} and Corollary \ref{cor.3.2},
 respectively.
 $\Box$
 \smallskip
 \end{pr}

 \begin{rem}
 \label{rem.3.4}
 Assume that $g\in{\cal G}^*$. The additional assumption $g\in\cal{I}$
 is neither necessary nor sufficient for the  EMS
 $\{g(k)\}_{k=1}^{\infty}$ to arise from an absolutely
 continuous
 \smallskip
 r.v.:

 \noindent
 (a) Consider the r.v.\ $X$ with density
 $f(x)=\frac{1}{2}I(-2<x<-1)+\frac{1}{2}I(1<x<2)$ so that
 $\mu_k=\mu_k(X)=2(k/(k+1)-2^{-k})$. Hence, $\mu_1=0$,
 $\lambda=\mu_2-\mu_1=5/6$, and from (\ref{3.7}) we see that
 $F_Y=\frac{3}{5} F_1+\frac{2}{5} F_2$, where $F_1$
 is the degenerate d.f.\ at $\log 2$ and the d.f.\ $F_2$ has
 density $f_2(y)=6e^{-2y}(1-e^{-y})$, $y>0$.
 Since
 $\mu\left(\{\log 2\}\right)=\lambda\Pr(Y=\log2)=\frac{1}{2}$, the function
 $g(x)=2(x/(x+1)-2^{-x})=G_{s}(\mu)(x)\in \cal{G}$
 has a non absolutely continuous canonical measure $\mu$. Thus,
 $g\notin \cal{I}$.
 \smallskip
 \\
 (b) For $h_1(y)=I(0<y<1)$ and $s(y)=1$ (\ref{3.1b}) yields
 $I_s(h_1)(x)=g(x)=1-\big(1-e^{-x}\big)/x$, $x\geq 1$. It is easily checked
 that the particular EMS $\{g(k)\}_{k=1}^{\infty}$ corresponds to the d.f.\ $F$
 with inverse $F^{-1}(u)=(1+\log u)I(e^{-1}<u<1)$. However, this
 $F$ does not have a density, since it assigns probability $1-e^{-1}$
 at the point zero.
 \end{rem}

 \begin{exam}
 \label{exam.3.1}
 Let $\mu_k=k^\theta$, $0<\theta<1$, and  define
 $g(x)=x^\theta$, $x\geq 1$.
 The representation (\ref{3.1b})
 follows from
 \begin{eqnarray*}
 x^\theta = \int_{0}^x \frac{\theta}{t^{1-\theta}} dt
 =\frac{\theta}{\Gamma(1-\theta)} \int_{0}^x \int_{0}^{\infty}
 y^{-\theta} e^{-ty}
 dy dt
 =\frac{\theta}{\Gamma(1-\theta)}  \int_{0}^{\infty}
 y^{-\theta} \int_{0}^x e^{-ty} dt dy,
 \end{eqnarray*}
 where the change in the order of integration is justified by
 Tonelli's theorem.
 Therefore,
 \be
 \label{3.18}
 x^\theta=\frac{\theta}{\Gamma(1-\theta)}  \int_{0}^{\infty}
 \frac{1-e^{-xy}}{y^{1+\theta}}  dy, \ \ \ x\geq 0, \ \ 0<\theta<1.
 \ee
 Thus, (\ref{3.1b}) is satisfied with
 $h_1(y)=\beta_\theta y^{-1-\theta}$, where
 $\beta_\theta=\theta/\Gamma(1-\theta)>0$, and $s(y)=1$
 (note that (\ref{3.7}) suggests using a different
 function $s$, namely,
 $\widetilde{s}(y)=e^{-y}+(e^{-y}-e^{-2y})/(2^{\theta}-1)$;
 hence, $s$
 in the representations (\ref{3.1b}) or (\ref{3.2})
 need not be unique).
 Since (\ref{3.1a}) is obviously fulfilled,
 Corollary \ref{cor.3.3} shows that the
 sequence $k^\theta$ is an
 EMS. More precisely, Theorem \ref{theo.3.2} shows that
 the
 particular EMS, $k^{\theta}$, corresponds to the
 r.v.\ $X\in\cal{F}$
 with distribution inverse $G$ given by
 (\ref{3.15}) (with $h_1(y)=\beta_\theta y^{-1-\theta}$), that is,
 \[
 F^{-1}(u)=G(u)=\frac{\theta}{\Gamma(1-\theta)}
 \int \frac{(-\log u)^{-1-\theta}}{u}du
 =\frac{(-\log u)^{-\theta}}{\Gamma(1-\theta)}+C.
 \]
 Since $\mu_1=\int_{0}^1 F^{-1}(u)du=1$ we find $C=0$ and the parent d.f.\
 admits the explicit formula
 $F(x)=\exp\big(-\lambda x^{-1/\theta}\big)$, $x>0$, where
 $\lambda=\Gamma(1-\theta)^{-1/\theta}>0$; thus,
 $1/X$ is Weibull.
 Moreover, it is evident from Theorem \ref{theo.3.2} and
 (\ref{3.18}) that
 $\big\{(k+c)^\theta\big\}_{k=1}^{\infty}$
 is an EMS for every $c\in[-1,\infty)$ and $\theta\in(0,1)$, and
 the corresponding functions in the representation
 (\ref{3.1b}) are $h_1(y)=\beta_\theta e^{-c y}/y^{1+\theta}$
 and $s(y)=e^{cy}$.
 \end{exam}

 \begin{exam}
 \label{exam.3.2}
 Let $\mu_k=\log k$ and  define
 $g(x)=\log x$, $x\geq 1$.
 To see the representation (\ref{3.1b}) write
 (for $x>0$)
 \be
 \label{3.19}
 \log x = \int_{1}^x \frac{1}{t} dt
 =\int_{1}^x \int_{0}^{\infty}
 e^{-ty}
 dy dt
 =
 \int_{0}^{\infty}
 \int_{1}^x e^{-ty} dt dy
 =\int_{0}^{\infty}\frac{e^{-y}-e^{-xy}}{y} dy
 \ee
 showing that $h_1(y)=1/y$ and $s(y)=e^{-y}$.
 Again (\ref{3.1a}) is obviously fulfilled
 and Corollary \ref{cor.3.3} shows that the sequence
 $\log k$ is an EMS. More precisely, (\ref{3.15}) yields
 $F^{-1}(u)=-\log(-\log u)+C$, $0<u<1$.
 By the substitution
 $y=-\log u$ we find
 \[
 \mu_1=\int_{0}^{1} F^{-1}(u)\ du
 =C-\int_{0}^\infty e^{-y}\log y \ dy=C+\gamma,
 \]
 where $\gamma$ is Euler's constant; see, e.g., Lagarias (2013),
 p.\ 535.
 Since
 $\mu_1=\log 1=0$,
 it follows that $C=-\gamma$ and $F(x)=\exp(-e^{-(x+\gamma)})$ is an
 extreme-value (Gumbel) distribution. Furthermore,
 Theorem \ref{theo.3.2} and
 (\ref{3.19}) enable us to verify that
 $\big\{\log(k+c)\big\}_{k=1}^\infty$
 is an EMS for every $c\in(-1,\infty)$;
 the corresponding functions in the representation
 (\ref{3.1b}) are $h_1(y)= e^{-c y}/y$
 and $s(y)=e^{(c-1) y}$.
 \end{exam}

 \begin{exam}
 \label{exam.3.3}
 The harmonic number function was defined by Euler as
 \be
 \label{3.20}
 H(x)=\int_{0}^{1} \frac{1-u^x}{1-u} du =
 \int_{0}^{\infty}\frac{e^{-y}}{1-e^{-y}}\big(1-e^{-xy}\big) dy, \ \ \ x>-1;
 \ee
 see Lagarias (2013), p.\ 532. It satisfies
 \[
 H(0)=0, \ \ H(n)=1+\frac{1}{2}+\cdots+\frac{1}{n} \ \ (n=1,2,\ldots),
 \ \ \
 H(x)=H(x-1)+\frac{1}{x}, \ \  x>0.
 \]
 From Theorem \ref{theo.3.2} we conclude  that for every
 $c\in(-2,\infty)$, the sequence $\big\{H(k+c)\big\}_{k=1}^{\infty}$
 is an EMS from an absolutely continuous r.v.;
 indeed, (\ref{3.20}) shows that the function $g(x)=H(x+c)$ satisfies
 (\ref{3.1a}) and (\ref{3.1b}) with
 $h_1(y)=e^{-(c+1)y}/(1-e^{-y})$ and $s(y)=e^{cy}$.
 The standard Exponential
 corresponds to $c=0$ and the standard Logistic
 to $c=-1$; see
 Example \ref{exam.4.1}, below.
 The function $\psi(x)=\frac{d}{dx}\log\Gamma(x)=\Gamma'(x)/\Gamma(x)$
 admits a similar representation due to Gauss;
 see Lagarias (2013), p.\ 557. It follows that
 $\big\{\psi(k+c)\big\}_{k=1}^{\infty}$ is an EMS for $c>-1$. However,
 this fact is evident from the corresponding result
 for $H$, due to the relationship
 $\psi(x)+\gamma=H(x-1)$, $x>0$. Finally,
 the easily verified identity
 \[
 \mu_k:= 1+\frac{1}{2^\theta}+\cdots+\frac{1}{k^\theta} =
 \frac{1}{\Gamma(\theta)} \int_{0}^{\infty}
 \frac{y^{\theta-1} e^{-y}}{1-e^{-y}} \big(1-e^{-k y}\big) dy
 \ \ \ \ (\theta>0,  \ k=1,2,\ldots)
 \]
 shows that this $\mu_k$ is an EMS for every $\theta>0$
 (choose $h_1(y)=\Gamma(\theta)^{-1}y^{\theta-1}e^{-y}/(1-e^{-y})$ and
 $s(y)=1$ in (\ref{3.1b})).
 \end{exam}

 \begin{rem}
 \label{rem.3.1}
 It is known that the class of Bernstein functions is
 closed under composition; see Schilling {\it et al.} (2012), p.\ 28,
 Corollary 3.8. Therefore, the connection of EMS's to Bernstein functions
 (Theorem \ref{theo.3.1}) provides an additional tool
 in verifying that a given sequence is EMS. For instance,
 Example \ref{exam.3.1} with $c=-1$
 shows that $g_1(x):=(x-1)^\theta$ ($x\geq 1$, $0<\theta<1$) belongs to ${\cal I}$;
 thus, from
 Lemma \ref{lem.3.3}
 and Proposition \ref{prop.3.2},
 $B_1(x):=g_1(x+1)-g_1(1)=x^\theta$ ($x\geq 0$)
 is Bernstein. By the same reasoning, Example \ref{exam.3.2} (with $c=0$)
 shows that $B_2(x):=\log(x+1)$ ($x\geq 0$) is Bernstein and, hence,
 $\beta(x):=B_1(B_2(x))=(\log(x+1))^\theta$ ($x\geq 0$) is also a Bernstein
 function with LKR as in (\ref{3.6}). Observing that
 $a_0=\beta(0)=0$ and $a_1=\lim_{x\to\infty}\beta(x)/x=0$ we see that
 the LKR triplet of $\beta$ is of the form $(0,0;\nu)$, $\nu\neq 0$.
 Hence, Proposition \ref{prop.3.2} shows that for any $\theta\in(0,1]$,
 the function $g(x):=\beta(x-1)=(\log x)^\theta$ ($x\geq 1$)
 belongs to ${\cal G^*}$, and we conclude from Theorem 
 \ref{theo.3.1} that
 $(\log k)^\theta$ is an EMS. Notice that for any $\delta>0$,
 $(\log x)^{1+\delta}\notin {\cal G}$, since the second derivative
 changes its sign in the interval $(1,\infty)$; see Proposition
 \ref{prop.3.1}.
 \end{rem}

 \section{Sequences of expected ranges}
 \label{sec.4}
 \setcounter{equation}{0}

 Denote by $R_k(X)=X_{k:k}-X_{1:k}=\max_i X_i-\min_i X_i$
 the (sample) range based on $k$
 iid copies $X_1,\ldots,X_k$ of an r.v.\ $X$.
 In the present section we consider
 the similar question concerning expected ranges.
 That is, we want to decide whether a given sequence
 $\{\rho_k\}_{k=1}^\infty$ represents an {\it Expected Ranges
 Sequence} (ERS, for short), i.e.,
 whether there exists an integrable
 r.v.\
 $X$ with
 \[
 \E R_k(X)=\rho_k, \ \ \ \ k=1,2,\ldots \ .
 \]
 The following result is the range analogue of
 Theorem \ref{theo.1.2}.

 \begin{theo}
 \label{theo.4.1}
 A sequence $\{\rho_k\}_{k=1}^{\infty}$
 is an ERS of
 a non-degenerate integrable r.v.\
 if and only if the following three
 conditions are satisfied.
 \\
 {\rm (i)} $(-1)^{s+1} \Delta^s \rho_k>0$ for all $s\geq 1$ and $k\geq 1$.
 \\
 {\rm (ii)} $\rho_k=o(k)$ as $k\to\infty$.
 \\
 {\rm (iii)} $\rho_k=\sum_{j=1}^k (-1)^j {k\choose j} \rho_j$
 for all $k\geq 1$.
 \end{theo}
 \begin{pr}{Proof} The conditions (i)--(iii) are necessary. Indeed, if
 $\rho_k=\E R_k(X)$ for some integrable
 r.v.\ $X$
 with d.f.\ $F$ then we have
 \[
 \rho_k=\int_{-\infty}^\infty [1-F^k(x)-(1-F(x))^k] \ dx,  \ \ k\geq1.
 \]
 Therefore, for all $s\geq 1$, $k\geq 1$,
 \[
 (-1)^{s+1}\Delta^s \rho_k=\int_{-\infty}^\infty
 [F^k(x)(1-F(x))^s+F^s(x)(1-F(x))^k] \ dx>0,
 \]
 showing (i). With $F^{-1}(u)=\inf\{x:F(x)\geq u\}$, $0<u<1$,
 we can write
 \[
 \frac{\rho_k}{k}=\frac{\E R_k(X)}{k}=\int_{0}^{1} [u^{k-1}-(1-u)^{k-1}]
 F^{-1}(u) \ du \to 0, \ \ \mbox{as}\ \ k\to\infty,
 \]
 by dominated convergence; this verifies (ii). Finally,
 \begin{eqnarray*}
 \sum_{j=1}^k(-1)^j {k \choose j} \ \rho_j
 &  = &
 \int_{-\infty}^\infty \sum_{j=1}^k(-1)^j
 {k \choose j}[1-F^j(x)-(1-F(x))^j] \ dx
 \\
 &  = &
 \int_{-\infty}^\infty [1-F^k(x)-(1-F(x))^k] \ dx=\rho_k,
  \end{eqnarray*}
 which is (iii).

 Conversely, assume that (i)--(iii) hold, and consider
 the sequence $\mu_k=\frac{1}{2}\rho_k$.
 Obviously, the conditions (i)--(iii) of
 Theorem \ref{theo.1.2} are fulfilled by
 $\mu_k$. Hence, we can find an
 integrable r.v.\ $X$ such that
 $\E X_{k:k}=\frac{1}{2}\rho_k$ for all $k\geq 1$. Since, however,
 $\E X_{1:k}=-\sum_{j=1}^k (-1)^j {k\choose j} \E X_{j:j}$
 (for any integrable $X$), the condition (iii) yields
 $\E X_{1:k}=-\frac{1}{2}\rho_k$; thus,
 $\E [X_{k:k}-X_{1:k}]=\rho_k$,
 and the proof is complete.
 $\Box$
 \end{pr}

 \begin{rem}
 \label{rem.4.1}
 (a) Condition (iii) implies $\rho_1=0$ (trivial)
 and $\rho_3=\frac{3}{2}\rho_2$. Condition (i) shows that
 $0=\rho_1<\rho_2<\cdots$.
 \smallskip
 \\
 (b) The random variable $X$, constructed in the sufficiency
 proof of Theorem
 \ref{theo.4.1}, is symmetric, i.e., $X\law -X$ (where $\law$ means
 equality in distribution). To see this, let $Y_i=-X_i$
 with $X_i$ being iid copies of $X$ used in the proof. Then
 $\E Y_{k:k}=\E\max\{-X_1,\ldots,-X_k\}=
 -\E\min \{X_1,\ldots,X_k\}
 =\frac{1}{2}\rho_k=\E X_{k:k}$ for all $k\geq 1$; thus,
 by the result of
 Hoeffding we see that $X$ and $Y$ have the same
 d.f.
 In fact, this is the {\it unique
 symmetric} r.v.\ having the given expected ranges.
 Indeed, if $Y$ is any symmetric r.v.\ with
 $\E R_k(Y)=\rho_k$ then, since $\E Y_{k:k}=-\E Y_{1:k}$
 (by symmetry), we should have $\rho_k=2\E Y_{k:k}$ for all
 $k\geq 1$.
 \smallskip
 \\
 (c) For any integrable $Y$ we can find a {\it symmetric}
 integrable $X$
 with the same expected ranges. Indeed, if $\rho_k=\E R_k(Y)$
 for arbitrary $Y$ (not necessarily symmetric), then the sequence
 $\rho_k$ satisfies the conditions (i)--(iii) of Theorem
 \ref{theo.4.1}.
 Thus, based on these values $\rho_k$, we can construct $X$
 as in the necessity proof, and this $X$ is symmetric.
 This fact
 seems to be quite surprising at a first glance.
 However, we observe that
 a d.f.\ $F$ is symmetric (i.e., it
 corresponds to a symmetric r.v.\ $X$)
 if and only if $F^{-1}(u)=-F^{-1}((1-u)+)$, $0<u<1$,
 where $F^{-1}(t+)$ denotes the right hand limit of $F^{-1}$
 at the point $t\in(0,1)$.
 Using this, it is easy to
 verify that the left continuous inverses of the
 d.f.'s
 of $X$ and $Y$ are related through
 \be
 \label{4.1}
 F_X^{-1}(u)=\frac{1}{2}\left[F_Y^{-1}(u)-F_Y^{-1}((1-u)+)\right], \ \
 0<u<1.
 \ee
 We conclude that
 the r.v.\ $X$, whose distribution inverse is
 defined by (\ref{4.1}),
 is the unique symmetric r.v.\ with the same
 expected ranges as $Y$.
 \end{rem}

 \begin{exam}
 \label{exam.4.1}
 It is well-known that the order statistics from the
 exponential distribution have means
 \[
 \E Y_{i:k}=\sum_{j=k-i+1}^{k} \frac{1}{j}, \ \ 1\leq i\leq k,
 \]
 and, therefore,
 \[
 \rho_k=\E R_k(Y)=\E [Y_{k:k}-Y_{1:k}] = 1+\frac{1}{2}+\cdots+\frac{1}{k-1}
 \ \ (\rho_1=0).
 \]
 From Theorem \ref{theo.4.1} and Remark \ref{rem.4.1} we know that
 there exists a
 unique symmetric r.v.\ $X$ with expected ranges $\rho_k$.
 Since $F^{-1}(u)=-\log(1-u)$, (\ref{4.1}) shows that
 \[
 F_X^{-1}(u)=\frac{1}{2} \log\left(\frac{u}{1-u}\right), \ \ 0<u<1,
 \]
 which corresponds to a Logistic r.v.\ with mean
 zero and variance $\frac{\pi^2}{12}$. This is in accordance
 with the recurrence relation
 $\mu_{k+1}=\frac{1}{k}+\mu_k$, satisfied by
 the expected maxima of the standard Logistic distribution
 (with mean zero and variance $\frac{\pi^2}{3}$), first
 obtained by Shah (1970); see also Arnold
 {\it et al.}\ (1992), p.\ 83.
 \end{exam}

 \begin{exam}
 \label{exam.4.2}
 The expected ranges of a Bernoulli$(p)$ r.v.\
 are $1-p^k-(1-p)^k$. The same expected ranges
 are obtained from a three-valued r.v.,
 assigning (equal) probabilities $\min\{p,1-p\}$ at $\pm \frac{1}{2}$,
 and the remaining mass $1-2\min\{p,1-p\}$ at zero.
 \end{exam}

 \begin{rem}
 \label{rem.4.2}
 If $Y$ is symmetric around its mean $\mu$ then, obviously,
 the symmetric r.v.\ with the same expected
 ranges is $X=Y-\mu$. In particular, if $Y$ is Uniform$(a,b)$ then
 $X$ is Uniform$\left(-\frac{1}{2}(b-a),\frac{1}{2}(b-a)\right)$; if $Y$
 is $N(\mu,\sigma^2)$
 then $X$ is $N(0,\sigma^2)$. However, it should be noted that
 there exist non-normal (non-uniform)
 r.v.'s
 with expected ranges like normal (uniform); see
 Arnold {\it et al.}\ (1992), pp.\ 145--146.
 To highlight the situation, assume that
 $X$ is $N(0,1)$ with density $\phi$, and let $\Phi$ be
 its d.f.\
 with inverse $\Phi^{-1}$.
 Let $0<\epsilon<\sqrt{2\pi}$
 and define $h(u)=\Phi^{-1}(u)+u(1-u)\epsilon$. Then,
 $h\in L^1(0,1)$ and
 $h'(u)=\frac{1}{\phi(\Phi^{-1}(u))}+(1-2u)\epsilon>0$
 for all $u\in(0,1)$. The fact that $h'(u)>0$ is obvious for
 $0<u\leq \frac{1}{2}$
 and it remains to verify that
 \[
 \epsilon<\frac{1}{(2u-1)\phi(\Phi^{-1}(u))}, \ \ \ \frac{1}{2}<u<1.
 \]
 This is indeed satisfied because
 \vspace*{-.5ex}
 \[
 \inf_{1/2<u<1}\left\{\frac{1}{(2u-1)\phi(\Phi^{-1}(u))}\right\}=
 \frac{1}{\sup_{1/2<u<1}\left\{(2u-1)\phi(\Phi^{-1}(u))\right\}}\geq
 \sqrt{2\pi}
  \vspace*{-.8ex}
 \]
 since
  \vspace*{-.8ex}
 \[
 \sup_{1/2<u<1}\left\{(2u-1)\phi(\Phi^{-1}(u))\right\}
 =
 \sup_{x>0}\left\{(2\Phi(x)-1)\phi(x)\right\}\leq
 \sup_{x>0}\phi(x)=\frac{1}{\sqrt{2\pi}}.
  \vspace*{-.5ex}
 \]
 Defining the r.v.\
 $Y=h(U)$, where $U$ is Uniform$(0,1)$,
 we see that $F_Y^{-1}=h$; thus, $Y$ is non-normal, and
  \vspace*{-.7ex}
 \begin{eqnarray*}
 \E R_{k}(Y)
 &=&
 \vspace*{-.7ex}
 k\int_{0}^1 (u^{k-1}-(1-u)^{k-1}) \Phi^{-1}(u) \ du
 \\
 &&
 \hspace{20ex}
 +k \epsilon
 \int_{0}^1 (u^{k-1}-(1-u)^{k-1})u(1-u) \ du
 \\
 &=&
 \vspace*{-.7ex}
  k\int_{0}^1 (u^{k-1}-(1-u)^{k-1}) \Phi^{-1}(u) \ du
  =\E R_k(X) \ \mbox{ for all }k\geq 1.
 \vspace*{-.7ex}
 \end{eqnarray*}
 Similar examples can be found for most
 r.v.'s.
 For example, a Uniform$(0,1)$ r.v.\
 $X$
 has the same expected ranges as a Beta$(1/2,1)$
 r.v.\
 $Y$ with density $f_Y(y)=(2\sqrt{y})^{-1}I(0<y<1)$.
 \end{rem}

 From Remark \ref{rem.4.2} it is clear that,
 in contrast to the expected maxima sequences,
 the sequences of expected ranges are far from
 characterizing the location family of the
 distribution.

 We summarize these facts in the following theorem.
 \begin{theo}
 \label{theo.4.2}
 {\rm (i)}
 A sequence $\{\rho_k\}_{k=1}^{\infty}$ represents the expected ranges
 of an
 integrable r.v.\ if and only if
 it represents the expected maxima of a symmetric {\rm(around zero)}
 integrable r.v.
 \\
 {\rm (ii)} For every integrable $Y$ there exists a unique
 symmetric integrable $X$ with the same expected ranges as $Y$;
 $X$ and $Y$ are related through {\rm (\ref{4.1})}.
 \\
 {\rm (iii)} The integrable r.v.'s $X$ and $Y$ have
 the same expected ranges if and only if the {\rm (generalized)}
 inverses $F_X^{-1}$
 and $F_Y^{-1}$ of their d.f.'s
 satisfy
 \be
 F_X^{-1}(u)-F_Y^{-1}(u)=F_X^{-1}((1-u)+)-F_Y^{-1}((1-u)+),
 \ 0<u<1,
 \label{4.2}
 \ee
 that is, if and only if the function $h(u)=F_X^{-1}(u)-F_Y^{-1}(u)$
 is symmetric around $\frac{1}{2}$ for almost all $u\in(0,1)$.
 \end{theo}
 \begin{pr}{Proof} (i) and (ii) are discussed in Remark
 \ref{rem.4.1}; note that the
 symmetric r.v.\ $X$ whose expected maxima
 are the expected ranges of $Y$ is given by
 (cf.\ (\ref{4.1}))
 \[
 F_X^{-1}(u)=F_Y^{-1}(u)-F_Y^{-1}((1-u)+), \ \ 0<u<1.
 \]
 To see (iii), assume first that $h:=F_X^{-1}-F_Y^{-1}$
 is almost everywhere symmetric around
 $\frac{1}{2}$. Then
 \[
  \E R_k(X)-
 \E R_k(Y)=
 k\int_0^{1}
 [u^{k-1}-(1-u)^{k-1}]
 h(u) \ du=0 \ \
 \mbox{ for all } \ k\geq 1,
 \]
 because the integrand,
 $g(u)=[u^{k-1}-(1-u)^{k-1}] h(u)$,
 is antisymmetric around $\frac{1}{2}$ (i.e.,
 $g(1-u)=-g(u)$ for almost all $u$).

 Conversely, $\E R_k(X)=\E R_k(Y)$ for all $k$ implies
 \[
 \int_0^{1}
 [u^{k-1}-(1-u)^{k-1}]
 [F_X^{-1}(u)-F_Y^{-1}(u)] \ du=\int_{0}^1 u ^{k-1}g(u) \ du =0 \ \
 \mbox{ for all } \ k\geq 1,
 \]
 where $g(u)=[F_X^{-1}(u)-F_Y^{-1}(u)]-[F_X^{-1}(1-u)-F_Y^{-1}(1-u)]$.
 Since $g\in L^1(0,1)$ and $\int_0^1 u^n g(u)du=0$ for $n=0,1,\ldots$,
 it follows that $g=0$ almost everywhere in
 $(0,1)$. This means that for almost all $u\in(0,1)$,
 \[
 F_X^{-1}(u)-F_Y^{-1}(u)=F_X^{-1}(1-u)-F_Y^{-1}(1-u),
 \]
 which, taking left limits to both sides, yields
 (\ref{4.2}).
 \smallskip
 $\Box$
 \end{pr}

 Therefore, every ERS is just a translation
 of an EMS from a symmetric r.v.\ (around its mean), and we can
 apply Theorem \ref{theo.3.1} to get the following
 characterization.
 \begin{theo}
 \label{theo.4.3}
 Let ${\cal{X}}_s$ be the class of non-degenerate, integrable
 r.v.'s that are symmetric around their means.
 A sequence $\{\mu_k\}_{k=1}^{\infty}$ is an EMS from an
 r.v.\ $X\in {\cal{X}}_s$
 if and only if it can be extended to a function
 $g=G_{s}(\mu)\in\cal{G}^*$ and, furthermore, the
 (unique) measure
 $\mu$ in the canonical form of $g$ satisfies
 \be
 \label{4.3}
 \mu\big( (0,y]\big) = \mu\Big(\Big[-\log(1-e^{-y}),\infty\Big)\Big),
 \ \ \ 0<y<\infty.
 \ee
 If such an extension $g$ exists,  it is unique
 (and it is given by (\ref{3.7})).
 \end{theo}
 \begin{pr}{Proof}
 Let $\mu_k=\mu_k(X)$ be the EMS of an r.v.\ $X\in {\cal X}_s$.
 By Theorem \ref{theo.3.1}, $\mu_k$ admits an extension
 $g=G_s(\mu)\in\cal{G}^*$.
 Also, $X-\mu_1$ is symmetric around $0$ and,
 according to Theorem \ref{theo.4.2}(i),
 $\rho_k=\mu_k-\mu_1$ is an ERS. In particular,
 $\rho_k=\mu_k-\mu_1$ satisfies condition (iii) of Theorem \ref{theo.4.1},
 i.e.,
 $(\mu_k-\mu_1)=\sum_{j=1}^k (-1)^j {k\choose j} (\mu_j-\mu_1)$,
 $k=1,2,\ldots$ \ .
 Substituting
 $\mu_j-\mu_1=g(j)-g(1)=
 \int_{(0,\infty)} h_0(y) \big(e^{-y}-e^{-jy}\big)d\mu(y)$
 ($j=1,2,\ldots,k$),
 we get
 \begin{eqnarray}
 \nonumber
 \hspace*{-6ex}
 \int_{(0,\infty)}
 \hspace{-1ex}
 h_0(y) \big(e^{-y}-e^{-ky}\big) d\mu(y)
 \hspace*{-1ex}& = &\hspace*{-1ex}
 \sum_{j=1}^k (-1)^j {k\choose j} \int_{(0,\infty)}
 \hspace{-1ex}
 h_0(y)
 \big(e^{-y}-e^{-jy}\big) \ d\mu(y)
 \\
 \label{4.4}
 \hspace*{-6ex}
  \hspace*{-1ex}&=&\hspace*{-1ex}
 \int_{(0,\infty)}
 \hspace{-1ex}
 h_0(y)
 \Big(1-e^{-y}-(1-e^{-y})^k\Big)d\mu(y),
 \ \ k=1,2,\ldots \ .
 \end{eqnarray}
 Consider the measure $\nu$ defined by $\nu\big((0,y]\big)
 =\mu\Big(\Big[-\log(1-e^{-y}),\infty\Big)\Big)$, $0<y<\infty$.
 Clearly, $\nu\neq 0$ is finite. Changing variables
 $y=-\log(1-e^{-w})$ in (\ref{4.4}), and since
 $h_0(-\log(1-e^{-w}))=h_0(w)$,
 $0<w<\infty$ (see (\ref{3.3})),
 we obtain
 \be
 \label{4.5}
 \int_{(0,\infty)}
  h_0(y) \big(e^{-y}-e^{-ky}\big) \ d\mu(y)
  =
 \int_{(0,\infty)}
  h_0(w) \big(e^{-w}-e^{-kw}\big) \ d\nu(w), \ \ k=1,2,\ldots \ .
 \ee
 Setting $s_0(y)=e^{-y}$ we see that the function
 $g_2:=G_{s_0}(\nu)\in \cal{G}^*$, and
 (\ref{4.5}) shows that $g(k)-g_2(k)=\mu_1$ (constant)
 for $k=1,2\ldots$; hence, $\mu=\nu$ (see Lemma \ref{lem.3.2}).
 Therefore, for all $y\in(0,\infty)$, $\mu\big((0,y]\big)=\nu\big((0,y]\big)
 =\mu\Big(\Big[-\log(1-e^{-y}),\infty\Big)\Big)$, $0<y<\infty$,
 and (\ref{4.3}) follows.

 Conversely, assume that there exists
 an extension  $g=G_s(\mu)\in{\cal G}^*$
 of $\mu_k$ with  $\mu$ satisfying (\ref{4.3}).
 Theorem \ref{theo.3.1} shows that $\mu_k$ is an EMS and,
 thus, $\rho_k=\mu_k-\mu_1$ is also an EMS. This means
 that the sequence $\rho_k$ satisfies the conditions (i) and (ii) of Theorem
 \ref{theo.4.1} (or of Theorem \ref{theo.1.2}).
 Moreover,
 \begin{eqnarray*}
 \sum_{j=1}^k (-1)^j {k \choose j} \rho_j
 &=&
 \sum_{j=1}^k (-1)^j {k\choose j} \int_{(0,\infty)}
 h_0(y)
 \big(e^{-y}-e^{-jy}\big) \ d\mu(y)
 \\
 &=&
 \int_{(0,\infty)}
 h_0(y)
 \Big(1-e^{-y}-(1-e^{-y})^k\Big) \ d\mu(y),
 \ \ k=1,2,\ldots \ .
 \end{eqnarray*}
 Substituting $y=-\log(1-e^{-w})$ in the last integral, and
 in view of (\ref{4.3}), it is easily seen that this integral equals
 $\rho_k$,
 and we conclude that the condition (iii)
 of Theorem \ref{theo.4.1} is also satisfied by $\rho_k$. Thus,
 $\rho_k$ is an ERS and, therefore, it is an EMS from
 a (unique) symmetric (around $0$) r.v.\ $Y$
 (see Theorem \ref{theo.4.2}(i)); that is, $\mu_k=\mu_1+\rho_k$
 is the EMS of $X=\mu_1+Y$, which is symmetric around its mean
 $\mu_1$.

 Uniqueness follows from Lemma \ref{lem.3.2}.
 $\Box$
 \end{pr}

 \begin{cor}
 \label{cor.4.1}
 A sequence $\{\rho_k\}_{k=1}^{\infty}$ is an ERS of
 a non-degenerate
 r.v.\ if and only if $\rho_1=0$ and there exists
 an extension $g=G_{s}(\mu)\in{\cal{G}^*}$
 of $\rho_k$
 such that the measure
 $\mu$ satisfies
 (\ref{4.3}).
 \end{cor}

 \begin{cor}
 \label{cor.4.2}
 Assume that
 the function $g$
 admits an integral representation
 of the form (\ref{3.1b}) with $h_1$
 satisfying (\ref{3.1a}); that is,
 $g=I_s(h_1)\in\cal{I}$. Then:
  \\
 (i) The sequence $\mu_k=g(k)$
 is an EMS
 of a symmetric (around its mean) non-degenerate r.v.\
 if and only if
 \be
 \label{4.6}
 h_1\left(-\log(1-e^{-y})\right)=(e^y-1)h_1(y)
 \ \ \
 \mbox{for almost all}
 \
 y\in(0,\infty).
 \ee
 (ii) The sequence $\rho_k=g(k)$
 is an ERS of a non-degenerate r.v.\ if and only if
 $\rho_1=0$ and (\ref{4.6}) is satisfied.
 \end{cor}
 \begin{pr}{Proof}
 The assumption on $g$ implies that $g\in\cal{I}\subseteq \cal{G}^*$ and
 thus, $g=G_s(\mu)$ for a unique $\mu\neq 0$ (see Lemmas
 \ref{lem.3.1}, \ref{lem.3.3} and Corollary \ref{cor.3.1}).
 From (\ref{3.1a}) we see that $\mu$ is absolutely continuous
 w.r.t.\ Lebesgue measure on $(0,\infty)$, with
 Radon-Nikodym derivative
 $h_\mu:=h_1/h_0$ (where $h_0(y)=e^{y}/(1-e^{-y})$; see (\ref{3.3})).
 Moreover,
 if $\nu$ is the measure defined by $\nu\big((0,y]\big)
 =\mu\Big(\Big[-\log(1-e^{-y}),\infty\Big)\Big)$, $0<y<\infty$,
 then $\nu$ is also absolutely continuous w.r.t.\ Lebesgue measure,
 since
 \[
 \nu\big((0,y]\big)=\mu\Big(\Big[-\log(1-e^{-y}),\infty\Big)\Big)
 =\int_{-\log(1-e^{-y})}^\infty h_{\mu}(x) \ dx,
 \ \ \
 0<y<\infty.
 \]
 From this expression it follows that a Radon-Nikodym derivative of $\nu$
 is given by
 \[
 h_{\nu}(y)
 :=\frac{d\nu(y)}{d y}
 =\frac{e^{-y}}{1-e^{-y}}h_{\mu}\Big(-\log(1-e^{-y})\Big),
 \ \ \ \
 0<y<\infty.
 \]
 Since $\mu=\nu$ if and only if $h_\mu=h_\nu$ a.e.\ in $(0,\infty)$,
 we conclude that (\ref{4.3}) is equivalent to (\ref{4.6}).
 The result follows from Theorems
 \ref{theo.4.3} and \ref{theo.4.2}(i).
 $\Box$
 \end{pr}

 \begin{exam}
 \label{exam.4.3}
 If $H$ is the harmonic number function,
 then $g(x):=H(x+c)=I_{s}(h_1)(x)$ ($c>-2$),
 where $h_1(y)=e^{-(c+1)y}/(1-e^{-y})$ and $s(y)=e^{cy}$;
 see (\ref{3.20}). It is easily seen that (\ref{4.6})
 reduces to $(e^y-1)^{c+1}=1$ a.e., and thus,
 it is satisfied if and only if $c=-1$. This shows that
 the only symmetric r.v.\ in this family is the Logistic,
 completing both Examples
 \ref{exam.3.3} and \ref{exam.4.1}.
 \end{exam}

 \begin{exam}
 \label{exam.4.4}
 For $g(x):=\log(x+c)=I_{s}(h_1)(x)$ ($c>-1$),
 $h_1(y)=e^{-cy}/y$ and $s(y)=e^{(c-1)y}$;
 see (\ref{3.19}). Hence,
 (\ref{4.6}) is written as
 $(e^y-1)^{c-1}=-\log(1-e^{-y})/y$ a.e.\
 and, obviously, this identity
 cannot be fulfilled (by any value of $c>-1$).
 Hence,
 all EMS's of Example \ref{exam.3.2}
 correspond to asymmetric r.v.'s.
 \end{exam}

 \begin{exam}
 \label{exam.4.5}
 For $g(x):=(x+c)^\theta=I_{s}(h_1)(x)$ ($c\geq -1$, $\theta\in(0,1)$),
 $h_1(y)=\beta_{\theta}e^{-cy}/y^{1+\theta}$ and $s(y)=e^{cy}$
 where $\beta_{\theta}>0$ is a constant;
 see (\ref{3.18}). Therefore, (\ref{4.6}) is now
 reduced to the identity
 $(e^y-1)^{c-1}=\big(-\log(1-e^{-y})/y\big)^{1+\theta}$ a.e.
 Obviously, this is impossible
 (for all values of $c\geq -1$ and $\theta\in(0,1)$).
 Hence,
 all EMS's of Example \ref{exam.3.1}
 correspond to asymmetric r.v.'s.
 \end{exam}

 \begin{exam}
 \label{exam.4.6}
 For $g(x):=1-1/(x+c)=I_{s}(h_1)(x)$ ($c> -1$),
 $h_1(y)=e^{-cy}$ and $s(y)=e^{(c-1)y}$.
 Therefore, (\ref{4.6}) is now
 reduced to the identity
 $(e^y-1)^{c-1}=1$ a.e.
 Obviously, this identity is satisfied if and only if
 $c=1$ (which corresponds to a standard uniform r.v.).
 Hence, $\{g(k)\}_{k=1}^{\infty}$ is an EMS for every $c>-1$
 (Theorem \ref{theo.3.2}),
 but the corresponding r.v.\ is asymmetric,
 unless $c=1$. Using (\ref{3.15}) it is recognized that
 $c=0$
 corresponds to the r.v.\ $1-Y$, with $Y$ being
 standard Exponential.
 \vspace{1em}
 \end{exam}

 \noindent
 {\bf Acknowledgements.}
 I would like to cordially thank an anonymous referee who
 provided a detailed review with insightful comments
 (corrected some mistakes),
 resulting to a great improvement of
 the presentation.
 Thanks are also due to
 Ch.A.\ Charalambides,
 V.\ Nestoridis and D.\ Gatzouras for helpful discussions.

 {
 
 }
 \end{document}